\documentclass[9pt,letterpaper,twoside]{rho-class/rho}

\usepackage{aas_macros}
\usepackage{siunitx}

\doctype{Preprint}
\title{New Way to Date Globular Clusters: Brown Dwarf Cooling Sequences}

\author[{\textasteriskcentered},a]{Roman Gerasimov}

\affil[a]{Department of Physics and Astronomy, University of Notre Dame, Nieuwland Science Hall, Notre Dame, 46556, Indiana, USA}

\dates{Accepted for publication in Astrophysics and Space Science}

\corres{\textsuperscript{\textasteriskcentered}Corresponding author: \href{mailto:rgerasim@nd.edu}{rgerasim@nd.edu} (R. Gerasimov).}

\journal{Roman Gerasimov}
\begin{abstract}
As the oldest building blocks of our Galaxy, globular clusters retain the archaeological footprint of the early stellar environments. Accurate absolute ages of globular clusters are required to interpret this ancient record. Existing dating techniques often produce precise but discordant ages, suggestive of systematic errors in excess of 1 Gyr. The James Webb Space Telescope (JWST) has unlocked a new dating method that leverages the cooling behavior of previously unobservable brown dwarf members. With a largely independent set of systematic errors, this new method provides a new consistency test for more established methodologies. I present a likelihood-based histogram-free method to derive globular cluster ages from multi-band JWST photometry of cluster members near and below the hydrogen-burning limit. By applying the method to a large set of simulated observations, I establish that formal age errors (i.e. errors based on measurement uncertainties alone) under 0.2 Gyr are attainable for nearby globular clusters. I also evaluate the significance of associated systematic effects, including the chemical heterogeneity of globular clusters (multiple populations), unresolved binary systems and uncertainties in brown dwarf cooling rates. As with other methods of age determination, systematic effects dominate the error budget (in selected cases, by over an order of magnitude), but may be reduced with more sophisticated analysis. Finally, I provide a lookup table for determining the number of observations, exposure times and temporal baselines required to estimate the age of a given cluster to a prescribed precision.
\end{abstract}


\definecolor{darkgreen}{rgb}{0.0, 0.5, 0.0}
\newcommand{\annotation}[1]{\textbf{\textcolor{magenta}{[#1]}}}
\newcommand{\revision}[1]{\textcolor{red}{#1}}
\newcommand{\revisionsecond}[1]{\textcolor{darkgreen}{#1}}

\renewcommand{\annotation}[1]{}
\renewcommand{\revision}[1]{#1}
\renewcommand{\revisionsecond}[1]{#1}

\begin{document}

    \maketitle
    \thispagestyle{firststyle}

%----------------------------------------------------------

\section{Introduction}\label{sec:introduction}

Globular clusters are massive \revision{($10^4-10^6\ \mathrm{M}_\odot$, \citealt{GC_masses_2,GC_masses})}, roughly spherical and compact (half-light radii $\lesssim5\ \mathrm{pc}$, \citealt{GC_harris}) conglomerations of stars that are found in galaxies of nearly all morphological types \citep{extragalactic_GCs_review_1,extragalactic_GCs_review_2,GCs_in_dwarfs_2,GCs_in_dwarfs_3}, \revision{including hosts of proto-globular clusters at high redshift} \citep{high_z_GC,high_z_GC_2,high_z_GC_3}. This work focuses exclusively on Milky Way globular clusters, of which some $150$ are currently known \citep{GC_harris}. A few of the Milky Way globular clusters are relatively young with ages near $6\ \mathrm{Gyr}$ (e.g., Palomar\,12, Whiting\,1, \citealt{young_GC_2,young_GC}); however, the majority are older than $12\ \mathrm{Gyr}$ \citep{dotter_survey,Valcin_3}, and some \revision{likely formed within $0.1-0.3\ \mathrm{Gyr}$ of the onset of star formation in the Galaxy (e.g., NGC\,6397 and NGC\,6752, as indicated by beryllium cosmochronology, \citealt{beryllium_dating,beryllium_dating_2})}.

While the origin of globular clusters has not been fully established (see the reviews in \citealt{GC_review,GC_formation_review}, and comparison of formation models in \citealt{GC_formation_models}), it is likely that the Milky Way hosts a mixture of clusters that formed \textit{in situ} \citep{insitu_GC_0,insitu_GC}, and clusters that were accreted from galactic mergers \citep{exsitu_GC,Sagittarius_GCs,sausage_GCs}. The two populations can be identified by distinct kinematics \citep{in_situ_vs_ex_situ_kinematics_2} and age-metallicity relationships \citep{insitu_vs_exsitu_AMR}, among other properties. The ages of accreted clusters constrain the masses of their original host galaxies due to the accelerated star formation in gas-rich environments \citep{gc_age_mass_relationship,GC_environments_through_ages}. The relationships between ages, chemistries and kinematics of accreted globular clusters retain an imprint of the assembly history of the Milky Way \citep{MW_assembly,GC_trace_assembly,GC_trace_assembly_2,GC_trace_assembly_4,GC_trace_assembly_3,wet_mergers}. Likewise, the equivalent relationships in the properties of \textit{in situ} globular clusters trace the internal evolutionary processes in the Galaxy, providing the history \revision{of chemical enrichment in the Galactic bulge \citep{bulge_chemical_enrichment}, an upper bound on the age of the Galactic bar \citep{GC_bar_formation}, and the timescale} of formation and spin-up of the Galactic disk \citep{GC_disk_spinup}. Accurate measurements of globular cluster ages are thus of great importance to galactic archaeology.

\annotation{The following part of the introduction has been entirely rewritten to address referee's concern about the lack of a sufficiently broad overview of the subject. The updated version provides more historical context, lists and compares all large GC age catalogs, and includes a more detailed discussion of associated systematic errors. It also presents a comprehensive list of GC age determinations from alternative dating methods.}

\annotation{To address referee's concern about my earlier use of anecdotal comparisons, I also added Tables~1 and 2 that provide a quantitative comparison of large age compilations from different methods.}

\revision{The oldest (see the early work in \citealt{first_MSTO_1,first_MSTO_2}), most commonly used and most robust method for dating globular clusters leverages the age sensitivity of the main sequence turnoff (MSTO). In its most basic form, the MSTO method extracts ages from the luminosity \citep{MSTO_V} or temperature \citep{MSTO_Teff_2,MSTO_Teff} of the bluest main sequence stars in the cluster. The obtained estimates are sensitive to errors in the adopted stellar models, and degenerate with the chemical composition of the cluster, interstellar reddening and distance. More modern implementations of the MSTO method include upper main sequence and post-main sequence stars in the fit to alleviate some of those degeneracies \citep{MSTO_full_CMD_fitting,MSTO_full_CMD_fitting_2}, or focus on specific morphological features of the color-magnitude diagram that are carefully chosen to reduce systematic errors. Notable examples of the latter approach include the $\Delta V^\mathrm{HB}_\mathrm{TO}$ method \citep{MSTO_delta_V_0,MSTO_delta_V,MSTO_delta_V_2} that utilizes the magnitude difference between the turnoff point and the horizontal branch, and the $\Delta(B-V)$ method \citep{delta_BV_method,delta_BV_method_2} that uses the color difference between the turnoff point and the base of the red giant branch. The $\Delta V^\mathrm{HB}_\mathrm{TO}$ parameter is less sensitive to the treatment of convection in stellar models, while $\Delta(B-V)$ is more suitable for globular clusters with underpopulated horizontal branches.}

\revision{The key advantage of the MSTO method is its universality, as it is based on photometric observations of comparatively luminous cluster members (as opposed to, e.g., white dwarfs or brown dwarfs). The most complete ($>50$ globular clusters) MSTO-based age compilations have been derived from Hubble Space Telescope (HST) photometry by \citet{relative_age_survey} ($64$ clusters, relative ages), \citet{dotter_survey} ($60$ clusters), \citet{GC_ages_2} ($55$ clusters), \citet{Wagner_catalog} ($69$ clusters), and \citet{Valcin_1,Valcin_3} ($69$ clusters). The median formal errors of MSTO-based age measurements can be as low as $0.4\ \mathrm{Gyr}$ \citep{GC_ages_2}, or even $0.1\ \mathrm{Gyr}$ \citep{Wagner_catalog}; however, the error budget is expected to be dominated by systematic effects, which are more challenging to estimate. \citet{modeling_errors} derived the ages of $22$ globular clusters while varying a large number of model parameters (convective properties, opacities, diffusion coefficients, nuclear rates, etc). They found the average uncertainty in inferred ages to be $\sim1.6\ \mathrm{Gyr}$. The treatment of convection in particular is likely to be the dominant contributor to the modeling error budget \citep{Valcin_2}. \citet{GC_ages_2} reduced the errors due to convection by only considering the part of the color-magnitude diagram near the turnoff (similarly to the $\Delta V^\mathrm{HB}_\mathrm{TO}$ method), but estimated systematic errors of $1.5-2\ \mathrm{Gyr}$ due to the uncertainties in distance and chemical abundances. More recently, \citet{Valcin_3} reported the average age error of $0.8\ \mathrm{Gyr}$ by incorporating the distances from Gaia \citep{gaia_distances_1,Gaia_EDR3} and allowing variations in the convective mixing length. This result is broadly in agreement with both the discrepancies between the available MSTO age compilations (see Tables~\ref{tab:intro}~and~\ref{tab:intro_2}) and the detailed analysis of the error budget in \citet{MSTO_error_budget}, who showed that MSTO-based age errors of $0.5-0.75\ \mathrm{Gyr}$ are attainable with the largest contributions from uncertainties in distance, interstellar reddening, chemical composition and convective parameters.}

% The most commonly used dating method extracts ages from the color-magnitude diagrams of globular clusters by taking advantage of their approximately coeval nature. Model isochrones at the appropriate chemical composition can be fitted directly to the main sequence turnoff and/or post-main sequence members (e.g., \citealt{matteo_NGC6397,michele_47Tuc,roman_47Tuc}). While the main sequence is not generally sensitive to age, it may be included in the fit to reduce errors in other parameters such as metallicity \citep{GC_ages_including_MS}. Alternatively, the models may be compared to specific features of the color-magnitude diagram that are deliberately chosen to minimize sensitivity to extraneous parameters and prevent over-fitting. Examples include the $\Delta(B-V)$ method \citep{delta_BV_method} that uses the color difference between the turnoff point and the onset of the red giant branch, and the $\Delta V^\mathrm{HB}_\mathrm{TO}$ method \citep{delta_HB_TO_method,GC_ages_2} that utilizes the magnitude difference between the turnoff point and the horizontal branch. A major downside of this family of dating techniques is the uncertain convective behavior of post-main sequence stars that introduces a partial degeneracy between the inferred ages of globular clusters and the convective parameters adopted in the models \citep{GC_ages_convection}.

\revision{While the MSTO method is likely to remain the standard globular cluster age diagnostic in the foreseeable future, alternative dating techniques with distinct systematic errors are necessary to derive tighter age constraints, calibrate stellar models and verify self-consistency.}

\revision{In addition to the morphology of the color-magnitude diagram leveraged by the MSTO method, the distribution of stellar luminosities (i.e., the luminosity function) may be used to probe the ages of globular clusters \citep{Bergbusch_LF_models}. Since stars with different initial masses evolve at different rates, the relative star counts in luminosity bins must change over time. This approach was suggested by \citet{Paczynski_LF_age} as a means to reduce the systematic errors associated with the treatment of convection, since stellar luminosities are largely insensitive to the convective mixing length. Furthermore, the most age-sensitive part of the luminosity function is the so-called \textit{subgiant break}, which is typically $\sim0.5$ magnitudes brighter than the turnoff point \citep{Zoccali_LF_ages}. Therefore, the luminosity function method is largely insensitive to the initial mass function and dynamical evolution of the globular cluster, as stars near the subgiant break span a narrow range of stellar masses. \citet{LF_age_distance_degeneracy} noted that a strategically placed luminosity bin at the transition between the red giant and subgiant branches would be far more sensitive to distance than to age, thereby breaking the degeneracy between these two parameters.}

\revision{Since convective properties and distance are prominent sources of error in the MSTO method, the luminosity function method has the potential to be more accurate (\citealt{Paust_LF_ages} found age errors as low as $0.5\ \mathrm{Gyr}$ including the distance uncertainty; also see a review of error budgets in \citealt{MSTO_vs_LF}). In practice, luminosity functions near and above the turnoff point serve as more direct probes of stellar evolution and are therefore more sensitive to modeling errors (see the reviews in \citealt{Vandenberg_review,Renzini_review,Salaris_review}). Furthermore, the luminosity function method is strongly affected by observational biases, including incomplete photometry, counting errors and blending.}

\revision{The largest homogeneous compilation of globular cluster ages determined from luminosity functions is found in \citet{Zoccali_LF_ages} (18 clusters); however, the ages were calculated to the nearest $\mathrm{Gyr}$ without formal errors. Additionally, the luminosity function method has been applied to $\sim10$ globular clusters in other studies, including NGC\,5466, NGC\,6229, NGC\,7006 \citep{Cohen_LF_ages}, M68 \citep{Jimenez_M68}, M13 \citep{Ratcliff_M13}, M5, M55 \citep{Jimenez_M5_M55}, M2, M3, M14 and M92 \citep{Paust_LF_ages}. Most of these estimates are based on outdated stellar models and are not directly comparable to modern MSTO measurements. The study in \citet{Paust_LF_ages}, however, is more recent. The comparison of luminosity function and MSTO-based ages in Tables~\ref{tab:intro}~and~\ref{tab:intro_2} suggests that luminosity function ages tend to be $1-1.5\ \mathrm{Gyr}$ older than MSTO ages in \citet{GC_ages_2,Valcin_3}, but are in agreement with MSTO ages in \citet{dotter_survey,Wagner_catalog,Valcin_1} within $\lesssim 0.5\ \mathrm{Gyr}$. It must be noted that this comparison is derived from a very small sample of just three globular clusters (M14 -- the globular cluster with the most precise age estimate in \citealt{Paust_LF_ages} -- is not included in any major compilation of MSTO-based age measurements; however, \citealt{M14_age} found the age of M14 to be close to that of M5, which suggests that the luminosity function-based age of M14 is likely overestimated by $\mathrm{1\ Gyr}$ or more).}

% This drop is largely insensitive to the adopted convective parameters \citep{Jimenez_LF_age}, thereby addressing the key weakness of color-magnitude diagram fitting. On the other hand, the luminosity function also depends on the mass function (i.e., the distribution of stellar masses) in the cluster that cannot be \textit{a priori} known due to extensive dynamical evolution \citep{GC_mass_functions}. The degeneracy between ages and mass functions can be alleviated by only considering cluster members well above the turnoff point, where the relevant mass range is very narrow \citep{Paczynski_LF_age}. However, discarding a large fraction of age-sensitive measurements would reduce the overall precision of the method.

\revision{The advent of deep space-based photometry of globular clusters with the HST allowed the first detection of white dwarf members \citep{first_WD_1,first_WD_2,first_WD_3,first_WD_4}. Unlike main sequence stars, white dwarfs are not supported by nuclear fusion, giving rise to long-term cooling. For realistic stellar lifetimes (e.g., \citealt{analytic_evolution}) and initial mass functions (e.g., \citealt{Kroupa_IMF}), the rate of white dwarf production is fastest in the early history of the globular cluster. For this reason, the faint (cool) end of the white dwarf cooling sequence is noticeably over-populated \citep{WD_end_overpopulation}. The age of the globular cluster may be estimated from the position of the white dwarf over-density in the luminosity function \citep{WD_LF_fitting} or the color-magnitude diagram \citep{WD_CMD_fitting}. This dating technique is less sensitive to the chemical composition of the cluster, since white dwarf atmospheres are expected to be metal-free in the absence of active accretion (\citealt{WD_accretion}, see also tentative signatures of accretion recently identified in NGC\,6397, \citealt{rolly_6397}). However, independent systematic errors are introduced by the uncertain cooling behavior of white dwarfs that depends on the properties and evolutionary history of the progenitor, chemical stratification, and conductive opacities, among other effects \citep{WD_uncertainties_1,WD_uncertainties_2,WD_uncertainties_3}.}

\revision{Due to the extremely faint luminosities of old white dwarfs, only a handful of nearby globular clusters have published white dwarf-based ages, including M4 \citep{WD_LF_fitting,WD_age_M4,WD_ages_6397_47Tuc_M4,WD_age_M4_2,WD_age_M4_3}, NGC\,6397 \citep{WD_CMD_fitting,47Tuc_WD_age,WD_ages_6397_47Tuc_M4,NGC_6397_WD_age,rolly_6397,WD_age_M4_3}, 47\,Tuc \citep{47Tuc_WD_age,47Tuc_WD_age_2,WD_ages_6397_47Tuc_M4,47Tuc_WD_age_3}, NGC\,6752 \citep{NGC6752_WD_age,NGC6752_WD_age_2} and $\omega$\,Cen \citep{omega_cen_WD_age}. Out of these measurements, I chose a sample of most recent age determinations for each globular cluster that are both absolute and consistent with a brief duration of star formation such that a single cluster age may be meaningfully defined. I then compared this sample to MSTO-based measurements in Tables~\ref{tab:intro}~and~\ref{tab:intro_2}. While white dwarf ages are discrepant by over $1\ \mathrm{Gyr}$ from the MSTO-based measurements in both \citet{Valcin_1} and \citet{Valcin_3}, it is noteworthy that the inclusion of Gaia distances in the latter eliminates the asymmetry in this discrepancy. On the other hand, white dwarf ages are in close agreement ($\lesssim 0.3\ \mathrm{Gyr}$) with MSTO measurements in \citet{GC_ages_2}.}

\revision{In special cases, the ages of globular clusters may be estimated by dating individual members. A common example involves matching evolutionary models to the masses and luminosities of detached eclipsing binary stars, as suggested in \citet{DEB_origin}. Since the first discovery of such members in globular clusters \citep{DEB_origin_2,DEB_origin_3,DEB_origin_4}, elipsing binary-based ages have been obtained for $\omega$\,Cen \citep{DEB_omega_cen}, 47\,Tuc \citep{eclipsing_binary_dating,DEB_47Tuc,DEB_47Tuc_2}, M4 \citep{DEB_M4}, M55 \citep{DEB_M55}, NGC\,6362 \citep{DEB_6362}, and NGC\,3201 \citep{DEB_3201,DEB_3201_2}. The most recent of these measurements are compared to other dating methods in Tables~\ref{tab:intro}~and~\ref{tab:intro_2}. Alternative approaches to dating individual members involve cosmochronology, e.g. by leveraging production of beryllium through interactions with cosmic rays \citep{beryllium_dating,beryllium_dating_2} or radioactive decay \citep{radioactive_dating_1,radioactive_dating_2}; however, too few globular clusters have published cosmochronological ages for a detailed analysis of associated systematic errors.}

\begin{table}[h]
\centering
\caption{Average \textit{absolute} differences (in $\mathrm{Gyr}$) in globular cluster ages adopted from different compilations of measurements}\label{tab:intro}%
\begin{tabular}{l lllllll}
\toprule
 & VB13 & D10 & WK17 & V20 & V25 & LF & WD  \\
\midrule
D10&$0.8\pm0.1$ &  &  &  &  &  &  \\
WK17&$1.0\pm0.1$ & $0.4\pm0.1$ &  &  &  &  &  \\
V20&$1.1\pm0.1$ & $0.7\pm0.2$ & $0.7\pm0.1$ &  &  &  &  \\
V25&$0.7\pm0.1$ & $0.7\pm0.1$ & $0.8\pm0.1$ & $1.1\pm0.1$ &  &  &  \\
LF&$1.1\pm0.7$ & $0.4\pm0.8$ & $0.2\pm0.6$ & $0.3\pm0.8$ & $1.3\pm0.7$ &  &  \\
WD&$0.3\pm0.3$ & $0.8\pm0.4$ & $0.9\pm0.3$ & $1.1\pm0.5$ & $1.4\pm0.4$ & $-$ &  \\
DEB&$0.6\pm0.2$ & $0.8\pm0.3$ & $1.6\pm0.2$ & $1.3\pm0.4$ & $1.1\pm0.4$ & $-$ & $-$ \\
\end{tabular}
\par\smallskip
\begin{minipage}{\linewidth}
\footnotesize
\raggedright
\noindent The averages shown in the table are mean values weighted by the formal errors of the age estimates added in quadrature. The errors in the averages were estimated by bootstrapping under the assumption of Gaussianity. The cells of the table that correspond to comparisons of measurement compilations with themselves, or cells that correspond to comparisons already included elsewhere in the table, are left blank. No averages were calculated for pairs of measurement compilations that have fewer than $3$ globular clusters in common, and the corresponding cells were filled with ``$-$''.

\medskip

\noindent VB13: MSTO measurements from \citet{GC_ages_2}; D10: MSTO measurements from \citet{dotter_survey}; WK17: MSTO measurements from \citet{Wagner_catalog}; V20: MSTO measurements from \citet{Valcin_1}; V25: MSTO measurements from \citet{Valcin_3}; LF: luminosity function measurements from \citet{Paust_LF_ages}; WD: most recent measurements of absolute ages based on white dwarf cooling sequences from \citet{WD_age_M4_2,NGC_6397_WD_age,47Tuc_WD_age_3,NGC6752_WD_age_2}; DEB: most recent measurements based on observations of detached eclipsing binary members from \citet{DEB_47Tuc_2,DEB_M4,DEB_M55,DEB_6362,DEB_3201_2}.
\end{minipage}
\end{table}

\begin{table}[h]
\centering
\caption{Average \textit{signed} differences (in $\mathrm{Gyr}$) in globular cluster ages adopted from different compilations of measurements}\label{tab:intro_2}%
\begin{tabular}{l lllllll}
\toprule
 & VB13 & D10 & WK17 & V20 & V25 & LF & WD  \\
\midrule
D10&$+0.8\pm0.1$ &  &  &  &  &  &  \\
WK17&$+1.0\pm0.1$ & $+0.3\pm0.1$ &  &  &  &  &  \\
V20&$+1.0\pm0.1$ & $+0.3\pm0.2$ & $+0.2\pm0.1$ &  &  &  &  \\
V25&$+0.3\pm0.1$ & $-0.5\pm0.1$ & $-0.6\pm0.1$ & $-0.8\pm0.1$ &  &  &  \\
LF&$+1.1\pm0.7$ & $+0.4\pm0.8$ & $+0.2\pm0.6$ & $+0.3\pm0.8$ & $+1.3\pm0.7$ &  &  \\
WD&$+0.2\pm0.3$ & $-0.6\pm0.4$ & $-0.9\pm0.3$ & $-1.1\pm0.5$ & $-0.2\pm0.4$ & $-$ &  \\
DEB&$-0.3\pm0.2$ & $-0.7\pm0.3$ & $-1.6\pm0.2$ & $-1.3\pm0.4$ & $-0.7\pm0.4$ & $-$ & $-$ \\
\end{tabular}
\par\smallskip
\begin{minipage}{\linewidth}
\footnotesize
\raggedright
\noindent The footnote of Table~\ref{tab:intro} applies to this table as well.

\medskip

The differences are shown as \textit{row} $-$ \textit{column}. For example, the value in the top-left cell of the table ($0.8\pm0.1\ \mathrm{Gyr}$) suggests that MSTO ages in \citet{dotter_survey} are older than the MSTO ages in \citet{GC_ages_2}, on average.
\end{minipage}
\end{table}

% When multiple dating techniques are applied to the same globular cluster, the inferred ages often disagree by over $1\ \mathrm{Gyr}$. For example, for the globular cluster M92, the available estimates include $11\pm1.5\ \mathrm{Gyr}$ (isochrone fitting, \citealt{M92_age_1}) and $12.75\pm0.25\ \mathrm{Gyr}$ (modified $\Delta V^\mathrm{HB}_\mathrm{TO}$ method, \citealt{GC_ages_2}). For NGC~6397: $11.5\pm\qty{0.5}{Gyr}$ (white dwarfs, \citealt{WD_CMD_fitting}), $12.6\pm\qty{0.7}{Gyr}$ (isochrone fitting, \citealt{matteo_NGC6397}) and $\gtrsim \qty{13.3}{Gyr}$ (beryllium abundance, \citealt{beryllium_dating}). In recent years, careful reevaluation of stellar models has led to significant progress in resolving these discrepancies. For example, the original white dwarf cooling age of 47\,Tuc ($9.9\pm0.7\ \mathrm{Gyr}$, \citealt{47Tuc_WD_age}) has been revised to $12\pm1\ \mathrm{Gyr}$ by \citet{47Tuc_WD_age_2}, bringing it much closer to the estimates from the $\Delta V^\mathrm{HB}_\mathrm{TO}$ method ($11.75\pm0.25\ \mathrm{Gyr}$, \citealt{GC_ages_2}) and an eclipsing binary member ($11.25\pm1\ \mathrm{Gyr}$, \citealt{eclipsing_binary_dating}). Nonetheless, the error budget is dominated by systematic effects in nearly all age measurements. The accuracy of absolute ages may be improved by comparing the measurements from a variety of dating techniques with independent systematics.

\begin{figure}[h]
\centering
\includegraphics[width=0.7\textwidth]{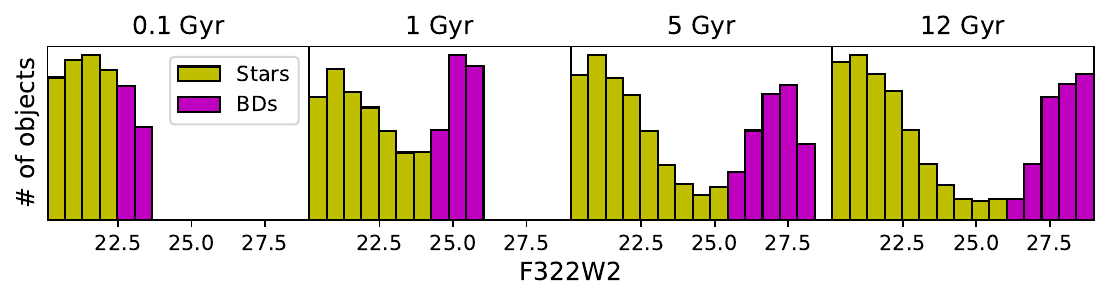}
\caption{Simulated luminosity function of a globular cluster with the metallicity and distance similar to those of 47\,Tuc. Individual subplots show the evolution of the luminosity function with cluster age (age increases from left to right). The horizontal axis displays the apparent magnitude in the \texttt{F322W2} band of the Near Infrared Camera on JWST. In the legend, ``BDs'' refers to brown dwarfs. The vertical axis is scaled linearly with an arbitrary scaling factor.}
\label{fig:intro}
\end{figure}

In this study, I propose a new method to date globular clusters using their brown dwarf cooling sequences. Brown dwarfs are substellar objects with masses below the minimum value required for hydrogen fusion (the \textit{hydrogen-burning limit}, $\sim 0.07-0.09\ \mathrm{M}_\odot$, \citealt{HBL_1,HBL_2,HBL_3,HBL_4}). The luminosity function of a typical globular cluster across the hydrogen-burning limit is shown in Figure~\ref{fig:intro} for multiple cluster ages. For the first $\sim0.1\ \mathrm{Gyr}$, the substellar component of the luminosity function (shown in magenta) evolves indistinguishably from low-mass stars above the hydrogen-burning limit. However, at later ages, the stellar members of the cluster begin to settle on the main sequence, while brown dwarfs continue cooling indefinitely. As a result, a gap emerges in the luminosity function (the \textit{stellar/substellar gap}, \citealt{adam_gap,Chabrier_gap}).

The detailed morphology of the stellar/substellar gap is strongly age dependent. \citet{Silvestri_gap} proposed using the number density of brown dwarfs near the bright edge of the gap as an age diagnostic. However, the sparse mass sampling adopted in that study (in particular, the use of a single stellar mass below the hydrogen-burning limit) obscures the continuous nature of this feature. In more finely sampled models (e.g., \citealt{Chabrier_gap,roman_omega_cen}), the bright edge of the gap emerges as a gradual transition between low-mass stars that have not fully settled on the main sequence and brown dwarfs that have not yet significantly cooled to fully separate from the main sequence. As a result, the onset of the gap is challenging to definitively identify in the color-magnitude diagram.

\citet{Caiazzo_gap} instead focused on the faint edge of the gap, populated by more evolved brown dwarfs, and suggested using the number density of objects in this regime as an age indicator. However, as I shall demonstrate in Section~\ref{sec:completeness}, this part of the luminosity function is expected to lie at magnitudes that are challenging to access in realistic observing strategies. Furthermore, more detailed theoretical models \citep{roman_47Tuc} and preliminary observational constraints \citep{michele_47Tuc} indicate that the faint edge of the gap is likely to be around $1\ \mathrm{mag}$ fainter than predicted in \citet{Caiazzo_gap}, further restricting its observability.

The approach presented here leverages the full age sensitivity of the luminosity function across the stellar/substellar gap. By forward modeling the entire luminosity function of the brown dwarf cooling sequence, my method incorporates contributions from objects at both edges of the gap as well as within it, without relying on specific morphological features that may be impossible to identify and measure due to intrinsically smooth transitions or incomplete photometric sampling.

\annotation{The paragraph below lists some examples of brown dwarf-based ages in other stellar associations, as requested by the referee.}

\revision{The cooling behavior of brown dwarfs is routinely employed to estimate the ages of open clusters and young stellar associations\footnote{When stellar populations are too young to have a prominent stellar/substellar gap, the star-brown dwarf transition can be located by observing the so-called \textit{lithium depletion boundary}, see \citet{LDB}.} \citep{OC_ages_1,OC_ages_2,OC_ages_5,OC_ages_3,OC_ages_4}, as well as broader stellar populations in the Milky Way \citep{Dino_BD_ages}.} However, due to the faint luminosities and infrared colors of brown dwarfs, this dating method could not be applied to globular clusters prior to the launch of the James Webb Space Telescope (JWST). At present, only one globular cluster has an age estimate derived from its brown dwarf cooling sequence. In \citet{SANDee}, the age of NGC~6397 was calculated as $13.4\pm3.3\ \mathrm{Gyr}$ based on a $2\ \mathrm{hr}$ exposure of the cluster with JWST. The large error is driven by the small sample size that comprised only three brown dwarfs with confirmed cluster membership. The pilot study in \citet{SANDee} was primarily limited by the lack of a second JWST observation of the same field, which is necessary to determine the proper motions of fainter objects in order to verify their cluster membership. However, it is expected that future 2-epoch JWST observations of nearby globular clusters may reveal brown dwarf members in far larger numbers \citep{roman_omega_cen,roman_47Tuc}.

This paper is organized as follows. Section~\ref{sec:simulation} describes the process for generating realistic simulated observations of nearby globular clusters with JWST. In Section~\ref{sec:analysis}, a likelihood-based method is presented for determining globular cluster ages from brown dwarf cooling sequences. The method is tested against simulated observations, and expected \revisionsecond{formal} errors are derived for a variety of observing strategies. \annotation{To address referee's concern about unaccounted systematic errors, the following sentence was added to explicitly emphasize that the analysis in Section 3 is only concerned with formal errors, while the systematic errors are considered in Section 4.} \revision{As with other dating techniques reviewed earlier, systematic errors are expected to dominate the error budget.} Section~\ref{sec:systematics} provides a detailed evaluation of key systematic errors associated with the new dating technique. In Section~\ref{sec:other_gcs}, the results are generalized to $30$ nearby globular clusters for which the substellar members may be observed with JWST. Section~\ref{sec:conclusion} concludes this study.

\section{Simulated observations}\label{sec:simulation}

This section details the procedure for generating realistic simulated observations of globular clusters with JWST. To estimate cluster ages from the brown dwarf cooling sequence, we are specifically interested in luminosity functions that may be inferred from the color-magnitude diagrams of members near and below the hydrogen-burning limit. Milky Way globular clusters display a wide variety of chemical compositions, mass functions and dynamical properties. Ideally, a separate set of tailored simulations is required for each globular cluster. However, the large number of globular clusters with potentially observable brown dwarf members makes an individualized approach impractical. Instead, I based the initial set of simulations on the known properties of the globular cluster 47\,Tuc. I then developed a scaling relationship that can be used to generalize the results of my analysis to other globular clusters (Section~\ref{sec:other_gcs}).

47\,Tuc is a prime target for deep JWST photometry, as it is both one of the nearest globular clusters ($4.45\pm0.12\ \mathrm{kpc}$, \citealt{47Tuc_distance}) and far more massive ($\sim10^6\ \mathrm{M}_\odot$, \citealt{GC_masses,GC_masses_2}) than the clusters at smaller distances. The large mass of 47\,Tuc allows an accurate reconstruction of its luminosity function due to the increased photometric sample. In fact, multiple tentative JWST detections of brown dwarf members in 47\,Tuc have been reported in the literature \citep{rolly_BD,marino_47tuc_BD,michele_47Tuc}.

\subsection{Observing strategy}

While the substellar members are far more sensitive to the age of the globular cluster than the main sequence, the observed sample must span a wide range of magnitudes that includes both low-mass stars and brown dwarfs. This is because stars just above the hydrogen-burning limit may not have fully settled on the main sequence (see the discussion of ``undecided'' objects in \citealt{SANDee}), and because low-mass stars constrain the mass function and other parameters of the cluster, which may be degenerate with age.

The imaging must be carried out in at least two photometric bands in order to differentiate the brown dwarf cooling sequence from the white dwarf cooling sequence. The two sequences typically span comparable magnitude ranges, but exhibit a large difference in color due to the much higher effective temperatures ($T_\mathrm{eff}$) of white dwarfs \citep{WD_models}. Inaccessibility of sufficiently precise photometric colors for brown dwarf candidates was one of the key factors that prevented the discovery of brown dwarfs in globular clusters with the HST \citep{BD_hunt_2}.

The Near Infrared Camera (NIRCam) on JWST allows simultaneous imaging of the field in two bands using its \textit{short wavelength} (SW, $0.6-\qty{2.3}{\micro\meter}$) and \textit{long wavelength} (LW, $2.4-\qty{5.0}{\micro\meter}$) channels. An ultra-wide filter is provided in both channels (\texttt{F150W2} in SW and \texttt{F322W2} in LW) that delivers the highest signal-to-noise ratio for faint objects. For this reason, I used the \texttt{F150W2}$-$\texttt{F322W2} filter combination in all simulated observations.

To avoid contamination of the inferred luminosity function by non-members and background galaxies, an accurate estimate of proper motion is required for each object included in the analysis. Therefore, at least two separate observations of the same field are required with a sufficiently long \textit{temporal baseline} (delay between observations). The specific choices of the exposure time and temporal baseline depend on the mass and distance to the globular cluster, the angular separation of the observed field from the center of the cluster, and the density of non-member contaminants at the position of the cluster in the sky. In my simulations, I considered two exposure times of $\sim1$ and $\sim10$ hours; and two temporal baselines of $2$ and $5$ years ($2\times2=4$ observing strategies in total). Note that the $5\ \mathrm{yr}$ baseline approximately corresponds to the expected remaining lifetime of the telescope at the time of writing \citep{JWST_lifetime}.

\subsection{Model isochrones}\label{sec:isochrones}

A key component of simulated color-magnitude diagrams is the adopted set of theoretical mass-luminosity relationships that connect the present-day mass function of the globular cluster to its observed luminosity function. The required models must reach the lowest brown dwarf masses observable in globular clusters with JWST, and allow for variations in light element abundances that have been observed in nearly all globular clusters (see Section~\ref{sec:systematics_mpops}). The stellar/substellar transition is characterized by cool atmospheres ($T_\mathrm{eff}\ll3000\ \mathrm{K}$) with dominant molecular opacity. For this reason, the effect of light element abundances on photometry is largely determined either by the distribution of oxygen abundance ($[\mathrm{O/Fe}]$) in the cluster (if the infrared spectrum is dominated by $\mathrm{H_2O}$ absorption) or carbon abundances ($[\mathrm{C/Fe}]$, if the infrared spectrum is dominated by $\mathrm{CH_4}$ absorption). The effect of other elements in the \texttt{F150W2}$-$\texttt{F322W2} color space is subdominant (see Section~\ref{sec:systematics_mpops}).

The majority of stellar model grids in the literature that reach into the brown dwarf regime are only available for solar (e.g., \citealt{Baraffe_isochrones,ATMO}) or metallicity-scaled solar (e.g., \citealt{BT-Settl,B06_models}) compositions. The notable exceptions are \texttt{LOWZ} \citep{LOWZ} and \texttt{Sonora} \citep{Sonora,ElfOwl} models that are available for multiple $\mathrm{C/O}$ abundance ratios, as well as \texttt{SANDee} models \citep{SAND,SANDee} that include a range of $\alpha$-enhancements with oxygen being one of the $\alpha$-elements. The lowest metallicity in the \texttt{Sonora} grid, $\mathrm{[Fe/H]}=-0.5$, is too metal-rich for the majority of globular clusters including 47\,Tuc \citep{GC_metallicities}. On the other hand, \texttt{LOWZ} models are only available for brown dwarfs ($T_\mathrm{eff}\leq 1600\ \mathrm{K}$) and do not capture the stellar/substellar transition that may extend to much higher temperatures. I therefore adopted the mass-luminosity relationships from the \texttt{SANDee} grid for the simulated observations.

\begin{figure}[h]
\centering
\includegraphics[width=0.7\textwidth]{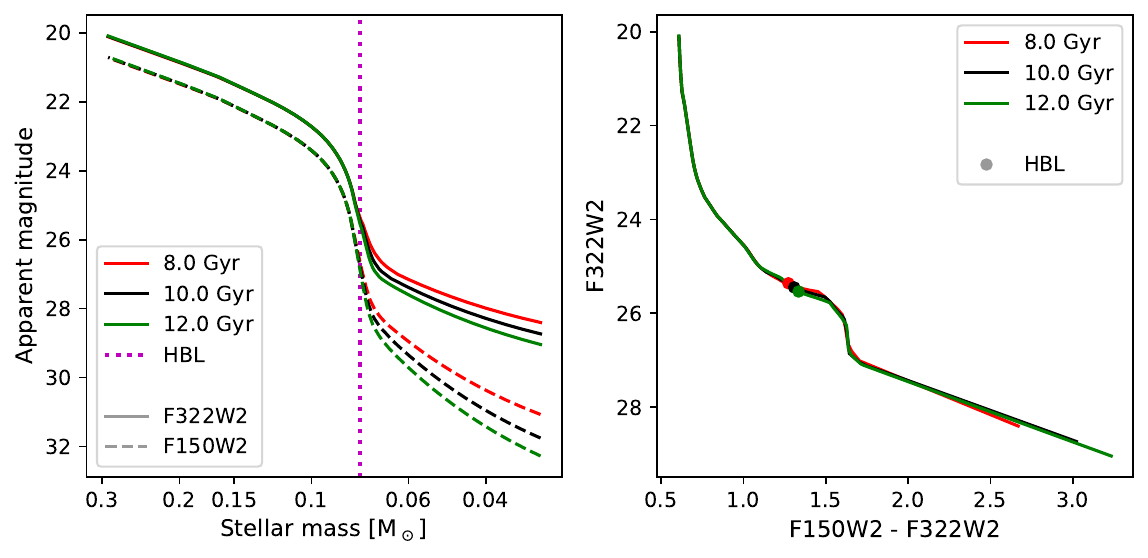}
\caption{Theoretical mass-luminosity relationships (\textit{left}) and color-magnitude diagrams (\textit{right}), extracted from the \texttt{SANDee} stellar models for the average chemical composition of 47\,Tuc. The ``true'' hydrogen-burning limit (HBL) of $0.077\ \mathrm{M}_\odot$ is shown for reference, as defined in \citet{SANDee}.}
\label{fig:MLR}
\end{figure}

For the initial set of simulations described in this section, I assumed that all members of the globular cluster have the same $[\mathrm{Fe/H}]$ and $[\mathrm{O/Fe}]\sim[\mathrm{\alpha/Fe}]$. I adopted $[\mathrm{Fe/H}]=-0.72$ from \citet{GC_harris}, and $[\mathrm{\alpha/Fe}]=0.17$ based on the mean oxygen abundance inferred from the spectroscopic measurements of giant members in \citet{nominal_C14,nominal_T14}. The effect of member-to-member variations in $[\mathrm{O/Fe}]$ is introduced in Section~\ref{sec:systematics_mpops}. The synthetic photometry from the \texttt{SANDee} models was interpolated to the adopted chemistry using linear Delaunay triangulation \citep{qhull}. In order to transform synthetic photometry into the observed plane, I used the optical reddening of $E(B-V)=0.04$ from \citet{roman_47Tuc}, the distance of $4.45\ \mathrm{kpc}$ from \citet{47Tuc_distance} and the extinction law from \citet{extinction}. The theoretical mass-luminosity relationships and color-magnitude diagrams are shown in Figure~\ref{fig:MLR} for a variety of cluster ages. Note that the color-magnitude diagram (right panel) is almost entirely insensitive to the age of the cluster, since the cooling tracks of brown dwarfs are generally aligned with the population isochrone. By contrast, the mass-luminosity relationship is sensitive to age below the hydrogen-burning limit.

\subsection{Mass function}

The mass function of the globular cluster determines the expected star/brown dwarf ratio in the observed field. To derive the age of the globular cluster from its luminosity function, the impact of the mass function and its partial degeneracy with age must be accounted for.

Mass functions are commonly represented by power laws due to the approximate scale invariance of star formation \citep{fractal_formation}, and the good agreement of power law models with the stellar mass distribution in the Galactic field \citep{Kroupa_IMF}. The mass functions of globular clusters display a wide variety of power law slopes, spanning both top-heavy and bottom-heavy mass distributions \citep{GC_mass_functions}. This diversity is likely driven by the dynamical evolution of globular clusters, as suggested by the measured relationships between the present-day mass function of a cluster, and its dynamical parameters such as central density or relaxation time \citep{MF_vs_environment,sollima_1}. It is therefore not unusual to infer different power law slopes from different observation fields within the same globular cluster (e.g., \citealt{GC_mass_functions,NGC6397_MF_1,SANDee}). The mass function below the hydrogen-burning limit is even more uncertain, since few mass measurements are available in this regime, and the sample of known brown dwarfs is heavily biased towards the solar neighborhood.

In this study, I model the mass functions ($\xi(M)$) of globular clusters as power laws with a change of slope (power law break) at $0.08\ \mathrm{M}_\odot$:

\begin{equation}
    \xi(M)\propto\begin{cases}
        M^{-\alpha_\mathrm{h}},& \text{if } M>0.08\ \mathrm{M}_\odot\\
        M^{-\alpha_\mathrm{l}},& \text{if } M\leq 0.08\ \mathrm{M}_\odot\\
    \end{cases}
    \label{eq:powerlaw}
\end{equation}

\noindent where $M$ is the stellar mass, and $\alpha_\mathrm{h}$ and $\alpha_\mathrm{l}$ are the high-mass and low-mass power law slopes, respectively. I chose the power law break of $0.08\ \mathrm{M}_\odot$ to match that of the ``universal'' mass function from \citet{Kroupa_IMF}. The break allows low-mass stars and brown dwarfs to have a distinct mass distribution from higher-mass members, e.g. due to contributions from alternative formation mechanisms that are unique to low-mass objects \citep{IMF_anomaly_2}. This potential slope discontinuity must be included in the simulations in order to constrain its contribution to the systematic error in the inferred cluster ages.

For the initial set of simulations, I set the high-mass and low-mass power law slopes to the ``universal'' values $\alpha_\mathrm{h}=1.3$ and $\alpha_\mathrm{l}=0.3$ from \citet{Kroupa_IMF}. The effects of other possible mass functions on the age estimates, including deviations from the power law approximation, are explored in Section~\ref{sec:systematics}.

\subsection{Measurement errors} \label{sec:errors}

Realistic errors in photometry and astrometry for a star with given NIRCam \texttt{F150W2} and \texttt{F322W2} magnitudes were calculated using the point-spread function (PSF) and the signal-noise relationship of the instrument from the JWST Exposure Time Calculator (ETC, \citealt{JWST_ETC}). For all ETC calculations, I used the \texttt{DEEP2} readout pattern, $20$ groups per integration, and $1$ or $9$ integrations per exposure for the total exposure times of $4100\ \mathrm{s}$ ($\sim1\ \mathrm{hr}$) and $37000\ \mathrm{s}$ ($\sim10\ \mathrm{hr}$), respectively. The sky background was set to ``low''.

\begin{figure}[h]
\centering
\includegraphics[width=0.7\textwidth]{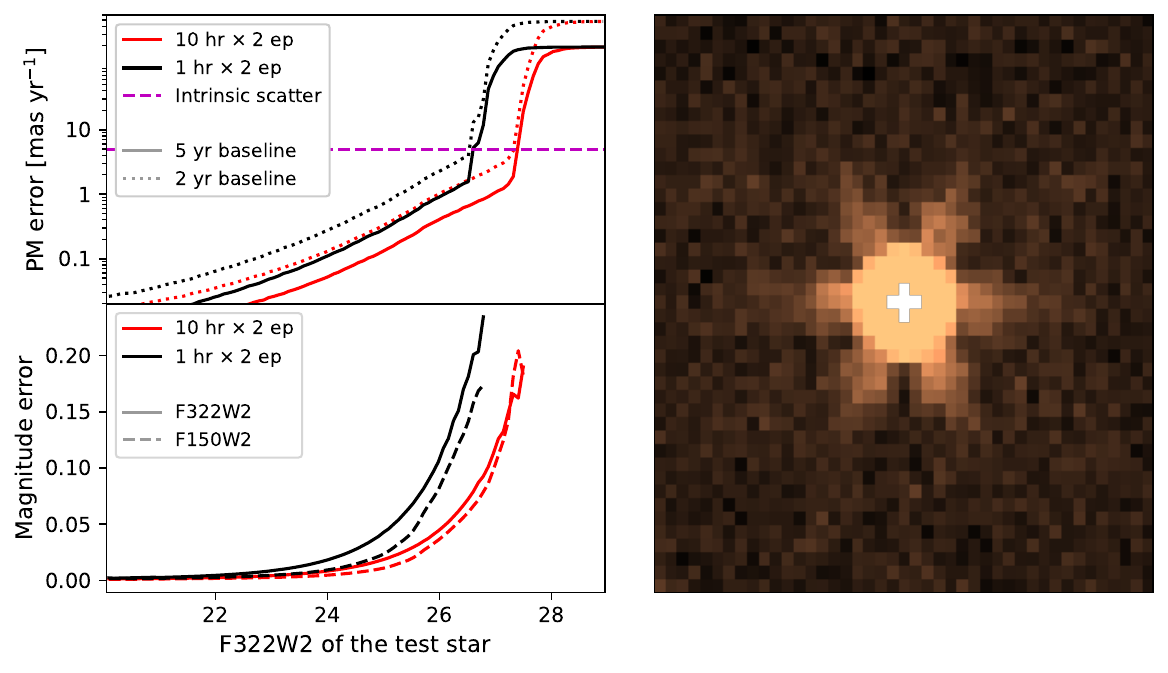}
\caption{\textit{Top left:} expected errors in the inferred proper motion (PM) of cluster members from NIRCam images as functions of the \texttt{F322W2} magnitude, for different observing strategies. The intrinsic scatter in proper motion among cluster members ($0.5\ \mathrm{mas\ yr}^{-1}$) is shown with a horizontal line. \textit{Bottom left:} expected errors in photometry for the two photometric bands considered in this study. Note that the expected errors in photometry (but not proper motion) are shown after the membership selection, as described in Section~\ref{sec:completeness}. Estimates are not available for the faintest test stars, since none of those stars passed the selection criteria. \textit{Right:} example of a simulated \texttt{F322W2} observation of a $20$th magnitude star with NIRCam. In the legends, times in hours refer to exposure times, and \textit{baseline} refers to the temporal baseline between the two observation epochs (\textit{ep}).}
\label{fig:errors}
\end{figure}

Simulated observations of stars were carried out as follows. First, the wavelength-dependent NIRCam PSF was averaged over all wavelengths, using the transmission profile of the photometric band (\texttt{F150W2} or \texttt{F322W2}) as weights. The wavelength-averaged PSF was then normalized to the simulated apparent magnitude of the star in the \texttt{VEGAMAG} system. I then added the sky background predicted by the ETC, and applied random Gaussian noise to both the object and the sky using the signal-noise relationships from the ETC. Finally, I masked pixels with signal above the saturation limit. The saturation limit was taken as the signal in the faintest pixel, flagged by the ETC in the image saturation map. An example of a simulated observation is shown in the right panel of Figure~\ref{fig:errors}.

I calculated measurement errors for $100$ test stars with evenly spaced \texttt{F322W2} magnitudes between $20$ and $29\ \mathrm{mag}$, and corresponding \texttt{F150W2} magnitudes according to the theoretical color-magnitude diagram described in Section~\ref{sec:isochrones} (also plotted in the right panel of Figure~\ref{fig:MLR}). Since the color-magnitude diagram is largely unaffected by cluster age, I used the arbitrarily chosen $11.5\ \mathrm{Gyr}$ isochrone for test stars, and assumed the calculated errors to be applicable to all ages. For each test star, I generated $300$ simulated observations, and assigned them into $150$ randomly chosen pairs that represent the two epochs of observations. The position of the star in the second epoch was offset by a random amount of proper motion, drawn from a normal distribution with the mean value of $(\mu_\alpha\cos(\delta),\mu_\delta)=(5.25,-2.53)\ \mathrm{mas}\ \mathrm{yr}^{-1}$, and the scatter of $0.5\ \mathrm{mas}\ \mathrm{yr}^{-1}$. These values were measured by \citet{47Tuc_pm,47tuc_pm_sigma} for 47\,Tuc. I have taken them to be representative of all globular clusters considered in this work. The calculations were repeated for two temporal baselines of $2$ and $5$ years.

The astrometric and photometric errors were extracted from the simulated observations by fitting the PSF model using the Goodman-Weare \citep{MCMC} Markov Chain Monte Carlo (MCMC) algorithm implemented in the \texttt{emcee} Python package \citep{emcee_2013}. The model PSF was constructed using the same procedure as the simulated observations but without applying noise to the image. PSF fitting was carried out on the simulated \texttt{F150W2} and \texttt{F322W2} images of the star simultaneously. The fitter was constrained to require the star to have the same position in both bands. The standard errors in magnitudes and positions were taken as the half-difference between the $84$th and $16$th percentiles ($\pm1$-sigma range) of the MCMC posteriors, and averaged over all $150$ trials. The results are plotted in the left panel of Figure~\ref{fig:errors}. Note that the errors in proper motion display a sudden rise near \texttt{F322W2} magnitude $~26-27$. This transition is indicative of the limiting magnitude of the observing strategy, beyond which the object is no longer detected in the image.

\subsection{Membership and completeness}\label{sec:completeness}

\textit{Photometric completeness} refers to the ratio of the number of detected cluster members at a given magnitude to the total number of members present in the observation field at that magnitude. Completeness decreases at fainter magnitudes, as more cluster members are overlooked either because they are too faint to be identified in the image, or because the errors in their photometry and astrometry are too large to differentiate them from white dwarf and non-member contaminants. The observed luminosity function must be corrected for the estimated completeness of the chosen observing strategy.

To estimate completeness from the simulated observations in Section~\ref{sec:errors}, I imposed two criteria that a star or brown dwarf must meet in order to be included in the analysis. First, the error in its \texttt{F150W2}$-$\texttt{F322W2} color must be lower than $0.25\ \mathrm{mag}$. This corresponds to half the color separation between white dwarfs and brown dwarfs in 47\,Tuc based on theoretical models \citep{WD_models} and previous JWST observations of the cluster \citep{michele_47Tuc}. Second, the measured proper motion of the candidate must be indicative of cluster membership with at least $95\%$ confidence.

The probability of a given star or brown dwarf being a member of the cluster can be estimated using the Bayes' theorem:

\begin{equation}
    P(\mathrm{mem}|\vec{\mu})=\frac{P(\vec{\mu}|\mathrm{mem}) P(\mathrm{mem})}{P(\vec{\mu}|\mathrm{mem})P(\mathrm{mem})+P(\vec{\mu}|\mathrm{mem'})P(\mathrm{mem'})}
    \label{eq:bayes}
\end{equation}

\noindent where $\mathrm{mem}$ stands for ``member'', $\mathrm{mem'}$ for ``non-member'' and $\vec{\mu}=(\mu_\alpha\cos(\delta),\mu_\delta)$ is the measured proper motion vector of the object. I took $P(\vec{\mu}|\mathrm{mem})$ to be a 2D Gaussian distribution, centered at the average proper motion of the cluster, and with the standard deviation given by the quadrature addition of the intrinsic proper motion scatter ($0.5\ \mathrm{mas\ yr}^{-1}$ along both axes, as defined in Section~\ref{sec:errors}) and the calculated errors in proper motion. For $P(\vec{\mu}|\mathrm{mem'})$, I took the proper motion distribution of field contaminants to be a 2D Gaussian centered at the origin with the scatter of $5\ \mathrm{mas\ yr}^{-1}$. These values are largely arbitrary, and were chosen to capture the common trend where the foreground distribution has a much wider scatter in proper motion than the cluster. In a real observation, the actual parameters of the field may be inferred e.g. by fitting a multi-component Gaussian distribution to the observed scatter in proper motion \citep{47tuc_pm_sigma}.

\begin{figure}[h]
\centering
\includegraphics[width=0.7\textwidth]{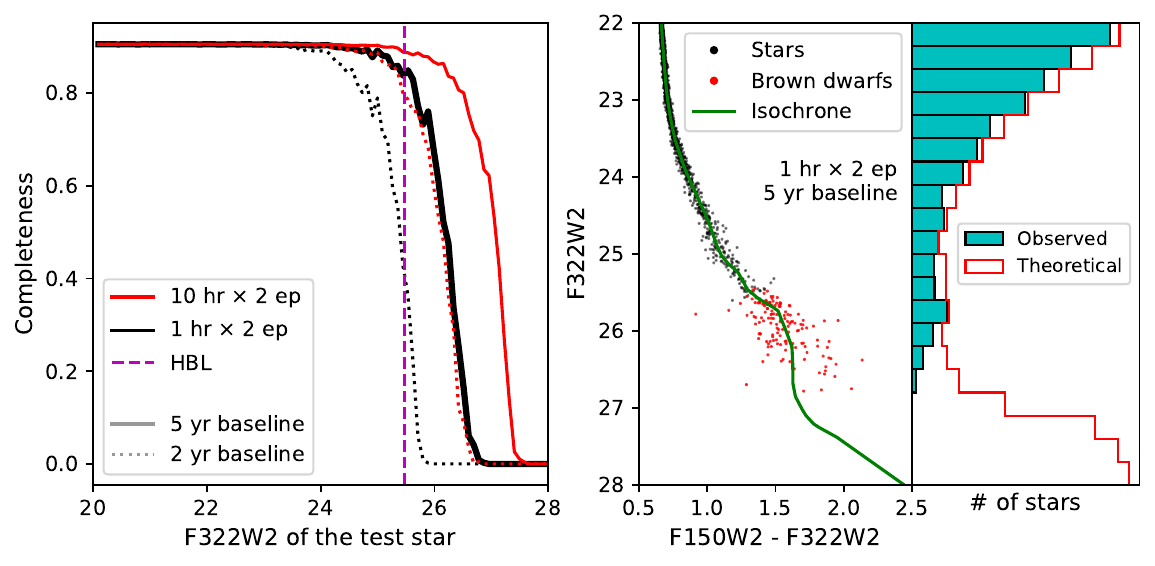}
\caption{\textit{Left:} estimated completeness of the observed luminosity function of 47\,Tuc for different observing strategies. In the legend, observing strategies are denoted in the same way as in Figure~\ref{fig:errors}. The magnitude of the hydrogen-burning limit (HBL) is shown for reference. \textit{Right:} simulated color-magnitude diagram (\textit{left subplot}) and luminosity function (\textit{right subplot}) for $2\times1\ \mathrm{hr}$ exposures taken $5$ years apart. The \textit{theoretical} luminosity function refers to the simulated luminosity function without measurement errors and completeness effects. The simulations in the right panel are shown for the age of $11.5\ \mathrm{Gyr}$. The luminosity function is shown on a linear scale with an arbitrary scaling factor.}
\label{fig:completeness}
\end{figure}

In the equation, $P(\mathrm{mem})=1-P(\mathrm{mem}')$ is the prior probability of cluster membership, determined by the ratio of members and contaminants in the observed field at a given magnitude. $P(\mathrm{mem}) \gg P(\mathrm{mem'})$ is expected at bright magnitudes, since within the cluster, its members are expected to vastly outnumber field stars. However, this may not apply to the magnitudes of the stellar/substellar transition, where the density of cluster members is greatly reduced by the stellar/substellar gap. In this study, I imposed no priors on cluster membership ($P(\mathrm{mem})=P(\mathrm{mem'})$), which may limit the accuracy of estimated completeness at faint magnitudes and introduce an additional systematic error in the inferred cluster age. This effect is discussed in Section~\ref{sec:systematics_modelling}.

For each test star, I estimated completeness as the fraction of simulated observations that passed the two aforementioned criteria. The result is plotted in the left panel of Figure~\ref{fig:completeness} for different observing strategies. The right panel of the figure shows an example of a simulated color-magnitude diagram and luminosity function of 47\,Tuc based on the theoretical mass-luminosity relationships, the adopted mass function and estimated photometric errors and completeness.

\annotation{To address referee's concern about the omission of photometric calibration errors in my analysis, the following added paragraph provides a reference to the JWST documentation that suggests that photometric calibration errors of $<0.01\ \mathrm{mag}$ are readily attainable. The estimated random errors in photometry below and near the hydrogen-burning limit are much smaller, so the effect of photometric calibration is likely inconsequential.}

\revision{I do not include the uncertainty in the absolute photometric calibration of NIRCam\footnote{\href{https://jwst-docs.stsci.edu/jwst-calibration-status/nircam-calibration-status/nircam-imaging-calibration-status}{https://jwst-docs.stsci.edu/jwst-calibration-status/nircam-calibration-status/nircam-imaging-calibration-status}} in my analysis, as it is expected to be much smaller than both random errors near the stellar/substellar gap (Figure~\ref{fig:errors}), and the effect of cluster age (Section~\ref{sec:analysis}).}

\section{Luminosity function analysis}\label{sec:analysis}

\annotation{To address referee's concern about unaccounted systematic errors, the following sentence was added to explicitly emphasize that the analysis in this section is only concerned with formal errors, while the systematic errors are considered in the following section.}

In this section, I describe the method to measure the age of a globular cluster from a JWST NIRCam observation of its brown dwarf cooling sequence. \revision{The aim of this section is to estimate the expected \textit{formal} errors associated with globular cluster ages, derived from brown dwarf cooling sequences. In this context, formal errors include photometric errors, completeness and algorithmic errors of the methodology described below. I note that the final error budget is likely dominated by systematic effects, which are discussed separately in Section~\ref{sec:systematics}.}

My method is based on fitting a model luminosity function to the observed magnitudes of cluster members. The standard approach to luminosity function fitting involves constructing a histogram from the observed magnitudes and comparing completeness-corrected counts in each bin to model predictions, e.g. \citet{omega_cen_IMF,roman_omega_cen,rolly_6397,SANDee}. However, histogram fitting has multiple major downsides: photometric errors may place measurements in incorrect bins, the results are sensitive to the histogram range and bin size, statistical errors in bin counts are not straightforward to estimate, and only one of the two photometric bands is typically used in the fitting process. In this study, I advocate against this approach, and propose obtaining the desired parameters through likelihood maximization instead.

The first step of the method is to construct a model luminosity function, $\phi(m)$, where $m$ is the observed magnitude of the star. The luminosity function is parameterized by the slopes of the mass function, $\alpha_\mathrm{h}$ and $\alpha_\mathrm{l}$ (Equation~\ref{eq:powerlaw}), and the age of the globular cluster. The luminosity function is related to the mass function by the derivative of the mass-luminosity relationship ($M(m)$):

\begin{equation}
    \phi(m;\ \mathrm{age},\alpha_\mathrm{h},\alpha_\mathrm{l})\propto\xi\left(M(m)\right) \left|\frac{dM(m)}{dm}\right|
    \label{eq:MF_to_LF}
\end{equation}

\begin{figure}[h]
\centering
\includegraphics[width=0.7\textwidth]{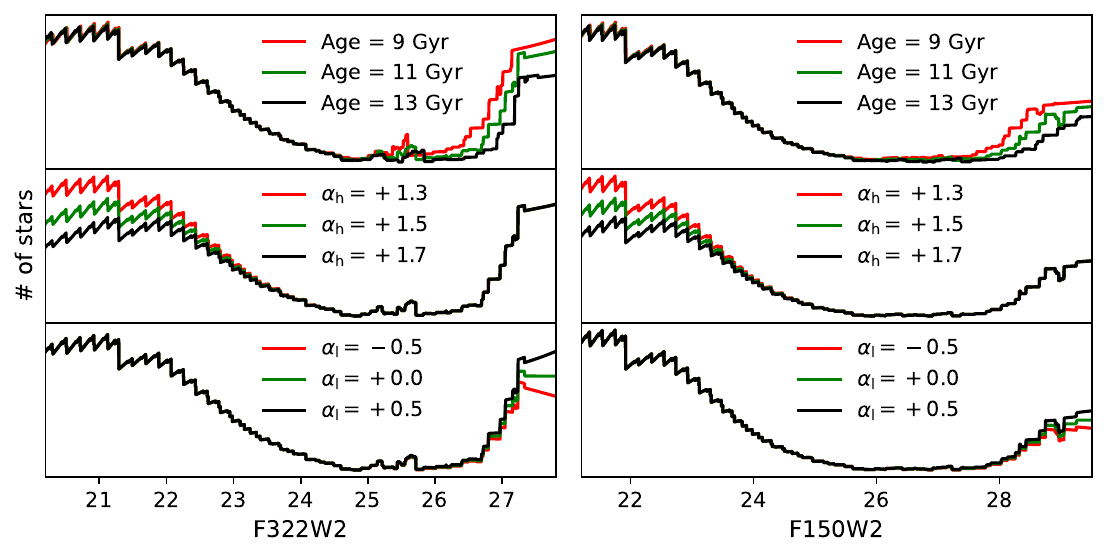}
\caption{Model luminosity functions for 47\,Tuc in \texttt{F322W2} (\textit{left}) and \texttt{F150W2} (\textit{right}) JWST NIRCam bands. The top, middle and bottom subplots show the effect of cluster age, high-mass power law slope and low-mass power law slope (see Equation~\ref{eq:powerlaw}) on the luminosity function, respectively. The vertical axis is linear and has arbitrary normalization.}
\label{fig:LF}
\end{figure}

The mass-luminosity relationship, $M(m)$, is obtained from theoretical stellar models as described in Section~\ref{sec:isochrones}. The relationship depends on the chemical composition of the cluster, interstellar reddening and distance. I computed the derivative of the mass-luminosity relationship numerically using first-order spline interpolation\footnote{In \texttt{SANDee} models, stellar masses are sampled using adaptive step-size control. While this approach can produce accurate synthetic color-magnitude diagrams, it is unideal for calculating derivatives due to the large variation in mass differences between adjacent stellar models. For this reason, the computed luminosity function shows noticeable numerical errors. For example, the ``sawtooth'' pattern at the bright magnitudes in Figure~\ref{fig:LF} corresponds to step-size adjustments in \texttt{SANDee}. I ignored this issue for the purposes of this study; however, a more consistent mass sampling is recommended for age measurements from real observations. Alternatively, the mass-luminosity relationship can be expressed as a linear combination of smooth functions prior to differentiation.}. Examples of model luminosity functions for 47\,Tuc and their dependence on age and the mass function are shown in Figure~\ref{fig:LF}.

The likelihood of compatibility ($\mathcal{L}$) between the observed magnitudes of cluster members, and a set of free parameters (age, $\alpha_\mathrm{h}$ and $\alpha_\mathrm{l}$) can be written as follows:

\begin{equation}
    \mathcal{L} = \prod_i  \int_{m_\mathrm{min}}^{+\infty}\left(\frac{\phi(m_i;\ \mathrm{age},\alpha_\mathrm{h},\alpha_\mathrm{l})\ \mathrm{comp}(m_i)}{N}\ \frac{1}{\sqrt{2\pi}\sigma_i}e^{-\frac{(x-m_i)^2}{2\sigma_i^2}}\right) dx
    \label{eq:likelihood}
\end{equation}

In the equation, $\mathrm{comp}(m)$ is the estimated completeness at magnitude $m$, while $m_i$ and $\sigma_i$ represent the full set of observed magnitudes and their errors, respectively. The set includes both \texttt{F150W2} and \texttt{F322W2} measurements. For practical purposes, it is convenient to compute $\phi(m)$ and $\mathrm{comp}(m)$ for only one of the bands (in my case, I chose \texttt{F322W2}), and convert all \texttt{F150W2} measurements into equivalent \texttt{F322W2} measurements using the theoretical color-magnitude diagram. $N$ is the normalization constant that makes the completeness-corrected luminosity function a valid probability density function:

\begin{equation}
    N=\int_{m_\mathrm{min}}^{+\infty}\phi(m;\ \mathrm{age},\alpha_\mathrm{h},\alpha_\mathrm{l})\ \mathrm{comp}(m)\ dm
    \label{eq:normalization}
\end{equation}

The lower limit of the integral, $m_\mathrm{min}$, is the brightest magnitude considered in the analysis. In a real observation, $m_\mathrm{min}$ is determined by the saturation limit of the instrument. In this study, I set $m_\mathrm{min}$ to the brightest magnitude within the range of \texttt{SANDee} models ($\approx 20.1$ in \texttt{F322W2}). I computed the integral in Equation~\ref{eq:likelihood} using the trapezoidal rule at $100$ evenly spaced points between $m_i-3\sigma_i$ and $m_i+3\sigma_i$. At the final step of the process, the logarithmic likelihood was maximized using the Nelder-Mead algorithm \citep{NelderMead}, implemented in the \texttt{scipy} Python package.

I applied the method described above to simulated observations of 47\,Tuc, generated as detailed in Section~\ref{sec:simulation}. For each simulated observation, the masses of observed stars and brown dwarfs were drawn from the ``universal'' mass function between the lower and upper mass limits of \texttt{SANDee} models at the adopted chemical composition of 47\,Tuc. The masses were then converted into synthetic \texttt{F150W2} and \texttt{F322W2} magnitudes by linearly interpolating the theoretical mass-luminosity relationship. Finally, a subset of generated stars was randomly removed from the sample according to the estimated photometric completeness, and Gaussian photometric errors were applied to all measurements.

I treated the total number of stars in the simulation as an independent parameter. Since the mass range of \texttt{SANDee} models does not have any physical significance, it is convenient to represent the total number of stars in the simulation indirectly, in terms of the number of stars within a certain range of magnitudes (henceforth referred to as ``richness'' of the observed field, $\mathcal{N}$). For simulated observations of 47\,Tuc, I chose the range between \texttt{F322W2} magnitudes $21$ and $22$, as this part of the color-magnitude diagram is well-populated, has high completeness and small photometric errors.

In a real observation, $\mathcal{N}$ is determined by the mass and dynamical properties of the cluster, as well as the angular separation between the observed field and the cluster center. In the NIRCam imaging program of 47\,Tuc analyzed in \citet{michele_47Tuc}, the combined field with multiple observation epochs spanned $\sim 12.4\ \mathrm{arcmin}^2$ and was located $\sim 8\ \mathrm{arcmin}$ from the cluster center. This area is comparable to the field of view of a single NIRCam image. This field yielded the richness of $\mathcal{N}\sim1500$, i.e. the total number of confirmed cluster members between \texttt{F322W2} magnitudes $21$ and $22$. I carried out simulated observations with the richness parameter set to integer multiples of this value between $\mathcal{N}\sim 1500$ and $\mathcal{N}\sim 15,000$. Each considered richness value represents a combination of multiple NIRCam observations of 47\,Tuc similar to the one in \citet{michele_47Tuc}.

\begin{figure}[h]
\centering
\includegraphics[width=0.7\textwidth]{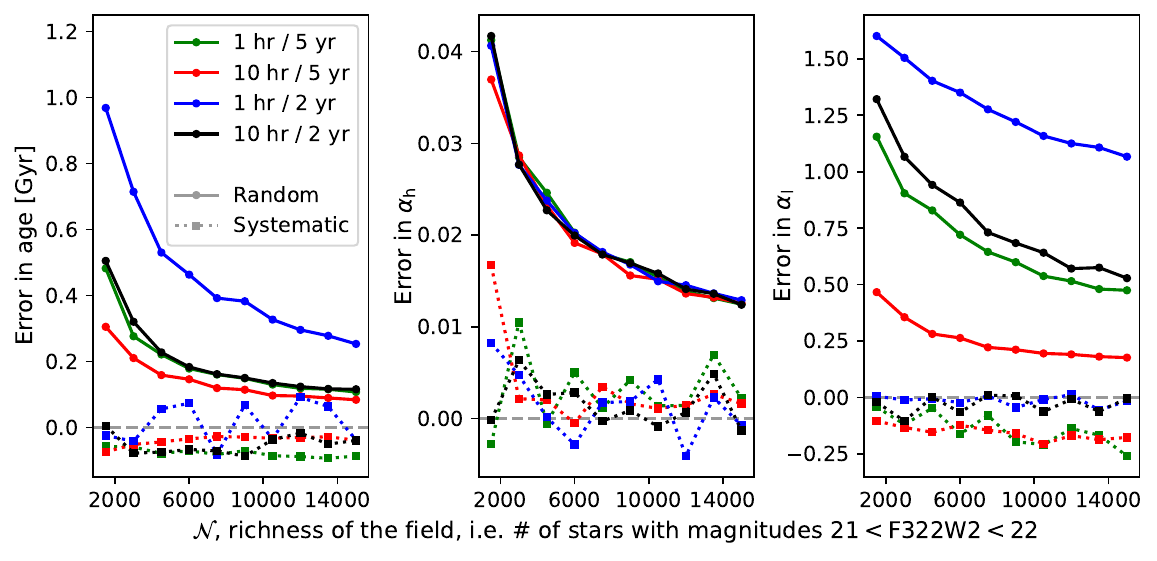}
\caption{\annotation{At face value, this figure appears to show ``systematic'' errors that are extremely small. However, in the context of Section 3, the term ``systematic errors'' refers to the systematic errors of the numerical algorithm, and not the physical systematic errors explored in Section 4. To make this clearer, I added a sentence to the figure caption and to the main text of Section 3 that emphasizes this further.} Expected random errors and systematic offsets in the measurements of the age and mass function parameters of 47\,Tuc from its brown dwarf cooling sequence. \revision{Note that the systematic offsets shown here include only algorithmic errors (see text). Physical systematic effects dominate the error budget, and are considered in Section~\ref{sec:systematics}.} The results are shown for 4 observing strategies with JWST NIRCam that correspond to different exposure times ($1\ \mathrm{hr}$ and $10\ \mathrm{hr}$) and temporal baselines between observation epochs ($2\ \mathrm{yr}$ and $5\ \mathrm{yr}$). The random errors are defined as standard deviations of the distribution of best-fit parameters from $1000$ simulated observations. Systematic offsets are the modes of these distributions. $\alpha_\mathrm{h}$ and $\alpha_\mathrm{l}$ are the slopes of the power-law mass function in Equation~\ref{eq:powerlaw}.}
\label{fig:random}
\end{figure}

For each considered value of $\mathcal{N}$, I generated $1000$ simulated observations per observing strategy, and calculated the best-fit age and mass function parameters for each simulation using likelihood maximization, as described above. I then estimated the expected formal errors in inferred age, $\alpha_\mathrm{h}$ and $\alpha_\mathrm{l}$ as the standard deviations of best-fit parameters across all simulations. I also calculated the expected systematic offset as the mode of the distribution, estimated using the Gaussian kernel density calculator in \texttt{scipy}. \revision{Note that this systematic offset only includes algorithmic errors and is expected to be very small. Physical systematic errors are much larger and are discussed in Section~\ref{sec:systematics}.} The results are shown in Figure~\ref{fig:random}.

The systematic offsets shown in Figure~\ref{fig:random} are primarily driven by asymmetric error bars in individual measurements, partial degeneracies between parameters and numerical artifacts. In most cases, these offsets are much smaller than the expected random errors, which suggests that the age determination algorithm is free of major flaws. For age measurements, two $10\ \mathrm{hr}$ exposures $2\ \mathrm{yr}$ apart produce nearly identical results to two $1\ \mathrm{hr}$ exposures $5\ \mathrm{yr}$ apart. The high-mass slope of the mass function ($\alpha_\mathrm{h}$) is primarily constrained by bright main sequence stars that are largely unaffected by completeness effects and photometric errors. For this reason, tight and nearly equivalent constraints are obtained from all observing strategies. By contrast, the range of observed stellar masses that contribute to $\alpha_\mathrm{l}$ is very small, resulting in weak constraints for most observing strategies.

\section{Systematic effects}\label{sec:systematics}

\annotation{The introduction to this section has been largely rewritten to provide a more structured overview of the systematic errors.}

\revision{Figure~\ref{fig:random} shows the expected errors in age and mass function parameters of 47\,Tuc as functions of richness, $\mathcal{N}$, for 4 different observing strategies of the cluster with JWST NIRCam. The key assumptions that were made in order to derive those estimates are \textit{(1)} \texttt{SANDee} models accurately capture the cooling behavior of metal-poor brown dwarfs, \textit{(2)} all cluster members have identical chemical abundances, (\textit{3}) said chemical abundances are known exactly, \textit{(4)} distance to the cluster is known exactly, \textit{(5)} absolute and differential reddening and their dependence on wavelength are correctly accounted for in synthetic photometry, \textit{(6)} the cluster contains no multiple star systems, \textit{(7)} the estimated photometric completeness is exact, \textit{(8)} the true age of 47\,Tuc is $11.5\ \mathrm{Gyr}$, and \textit{(9)} the true mass function of the cluster is given by the ``universal'' broken power law approximation in \citet{Kroupa_IMF}. All of these assumptions make contributions to the final error budget.}

\annotation{The following paragraph addresses the referee's concern about unaccounted systematic errors due to distance and reddening. While these errors are indeed excluded from my analysis, the paragraph below outlines why I anticipate their effects to be subdominant.}

\revision{The effects of distance \textit{(4)} and reddening \textit{(5)} are also some of the most prominent systematic errors in MSTO-based ages, as described in Section~\ref{sec:introduction}. For the MSTO method, these effects typically contribute between $0.5\ \mathrm{Gyr}$ \citep{MSTO_error_budget} and $2\ \mathrm{Gyr}$ \citep{GC_ages_2} to the error budget. The reddening error contribution must be much smaller for ages based on brown dwarf cooling sequences, because photometric colors are not directly utilized by this method, and because interstellar extinction is less important at infrared wavelengths. The error due to the uncertainty in distance is likely to be subdominant as well, because the effect of age on the width of the stellar/substellar gap (Fig.~\ref{fig:LF}) is much larger than the state-of-the-art uncertainties in distance moduli to nearby globular clusters derived from Gaia/HST astrometry \citep{gaia_distances_1}.}

\annotation{Another systematic effect brought up by the referee that was not included in my analysis is the dependence on the helium mass fraction. I justify this omission in the following paragraph by referring to a recent study by Scalco et al. 2025, to which I contributed stellar models with a wide range of helium content. To first order, the helium mass fraction has no effect on the luminosity function even though it does have a large effect on the mass-luminosity relationship.}

\revision{The effect of chemistry \textit{(2)}/\textit{(3)} may be broadly divided into three components: the effect of helium mass fraction ($\mathrm{Y}$), the effect of metallicity, and the effect of light element abundances. The helium mass fraction varies both from one globular cluster to another, and from member to member within the same cluster \citep{bedin_bifurcation,helium_in_GCs}. The value of $\mathrm{Y}$ affects stellar evolution by altering the rate of nuclear fusion and the mean molecular weight of the interior, among other effects. The corresponding offsets in the mass-luminosity relationship are expected to be significant for low-mass stars and brown dwarfs \citep{helium_1,helium_2,helium_3}. Fortunately, the error in age due to the uncertainty in the helium mass fraction is reduced by the fact that the morphology of the stellar/substellar gap is determined not by the mass-luminosity relationship directly, but by its derivative (Equation~\ref{eq:MF_to_LF}), which can be shown to be far less sensitive to $\mathrm{Y}$ \citep{michele_omegacen}. Nonetheless, the precise contribution to the error budget must be estimated with simulated observations at a variety of $\mathrm{Y}$ values. There are currently no publicly available grids of brown dwarf models that allow variations not only in $[\mathrm{Fe/H}]$ and $[\mathrm{\alpha/Fe}]$, but also $\mathrm{Y}$. As such, this analysis must be deferred to a future study. The effect of metallicity is considered in Section~\ref{sec:systematics_modelling}, and the effect of light elements is considered in Section~\ref{sec:systematics_mpops}.}

\annotation{The referee raised the concern that the effects of metallicity and the modeling errors are not accounted in my analysis. This is not entirely accurate, as both effects are explored simultaneously in Section 4.3. I added the paragraph below to introduce that subsection and explain why I believe the uncertainties in the treatment of model atmospheres in SANDee and the uncertainty in cluster metallicity have similar impacts on model isochrones.}

\revision{The modeling errors \textit{(1)} are particularly challenging to quantify. Multiple factors are expected to alleviate their contributions to the error budget. First, the dating technique proposed in this study does not directly utilize the colors of individual members. Unlike photometric colors, which are strongly influenced by stellar atmospheres, luminosities are largely determined by evolutionary models which are somewhat more robust (atmospheres contribute to evolution as well, but indirectly, see review in \citealt{Burrows_isochrones}). Second, the method only requires predicted magnitudes in ultra-wide bands, and not the detailed spectral energy distribution which is much harder to calculate accurately \citep{modelling_review}. These predictions are expected to be more robust in infrared bands that are dominated by more straightforward opacity sources such as $\mathrm{H_2O}$ absorption (as opposed to, e.g., hydride absorption or pressure-broadened alkali lines). Comparisons of \texttt{SANDee} models to observed spectra of brown dwarfs and low-mass stars can be found in \citet{Adam_T_index,Adam_MCMC_vs_AI}; however, it is not clear how the observed discrepancies propagate into age errors. \texttt{SANDee} models have been previously used to analyze the brown dwarf cooling sequence of NGC\,6397 in \citet{SANDee}. That study showed excellent agreement between the observed and predicted color-magnitude diagrams of brown dwarf members; however, a metallicity offset from the spectroscopically inferred value was required. Based on this result, it is perhaps reasonable to approximate modeling errors as offsets in metallicity between the simulated observations and the isochrones used in the analysis. This evaluation is carried out in Section~\ref{sec:systematics_modelling}.}

\revision{The effect of binary stars \textit{(6)} is investigated in Section~\ref{sec:systematics_binaries}. The effect of errors in the photometric completeness \textit{(7)} is simulated in Section~\ref{sec:systematics_modelling}. The effects of age and mass function \textit{(8)/(9)} are discussed in Section~\ref{sec:systematics_age_MF}.}

\begin{figure}[h]
\centering
\includegraphics[width=0.7\textwidth]{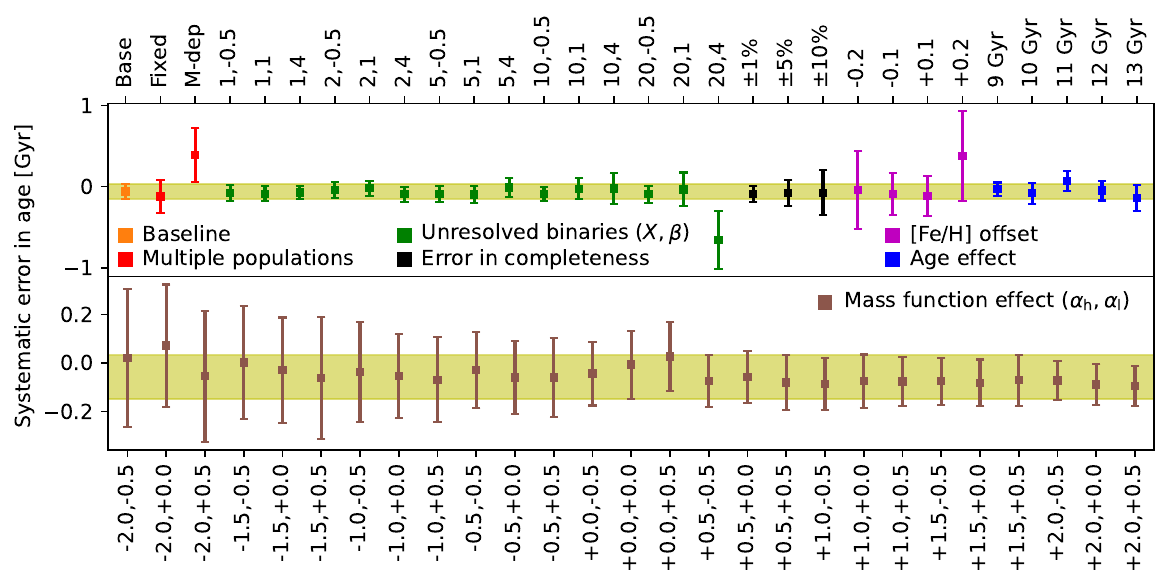}
\caption{Expected systematic effects in globular cluster ages inferred from their brown dwarf cooling sequences. Each effect is represented with a marker, centered at the modal offset from the true age induced by the systematic effect, and with error bars representing the expected spread in best-fit age due to the systematic effect. ``Base'' refers to the baseline case without any systematic effects. For reference, the baseline spread is also shown with yellow shading. Red markers represent the effect of multiple populations (Section~\ref{sec:systematics_mpops}) for the fixed and mass-dependent (M-dep) cases. Green markers represent the effect of unresolved binaries (Section~\ref{sec:systematics_binaries}), parameterized by the binary fraction $X$ (first number) and the slope of the companion mass distribution $\beta$ (second number). Black markers represent the effect of errors in the estimated photometric completeness (Section~\ref{sec:systematics_modelling}). Magenta markers show the effect of using a ``wrong'' mass-luminosity relationship in the analysis, represented here as an offset in the assumed metallicity of the cluster (Section~\ref{sec:systematics_modelling}). Blue markers show the expected modal offsets and scatter for different true ages of the cluster (Section~\ref{sec:systematics_age_MF}). Finally, brown markers do the same for various true mass functions, parameterized by $\alpha_\mathrm{h}$ and $\alpha_\mathrm{l}$ as in Equation~\ref{eq:powerlaw}.}
\label{fig:systematic}
\end{figure}

To quantify the contributions of major systematic effects to the error budget, I ran additional tests by generating a set of $200$ simulated observations for each considered effect, as described below. All simulations described in this section adopted two $1\ \mathrm{hr}$ exposures with a $5\ \mathrm{yr}$ temporal baseline, and had $\mathcal{N}\sim15,000$ -- the highest richness considered in this study. The high value of $\mathcal{N}$ in these tests was chosen to minimize random errors and isolate genuine systematic effects that persist even for very large sample sizes. The standard deviations of best-fit age in each set of simulated observations, and the modes of these values are shown in Figure~\ref{fig:systematic}. The figure also displays the baseline case, which corresponds to a set of simulated observations without any systematic effects, similar to the simulations in Section~\ref{sec:analysis}.

\revision{The results shown in Figure~\ref{fig:systematic} should be interpreted as follows. If a particular systematic effect increases the spread in best-fit ages across the set of test simulations, then the ratio of increased to baseline spreads should determine the scaling factor by which the formal errors must be adjusted. For example, a $10\%$ error in photometric completeness inflates the spread by approximately a factor of three. Therefore, the contribution of this systematic effect to the overall error budget is approximately twice as large as the formal error. On the other hand, a systematic effect that offsets the spread of best-fit ages is expected to have a constant contribution to the error budget regardless of the formal error. For example, Figure~\ref{fig:systematic} shows that an offset in the adopted metallicity of the cluster by $+0.2\ \mathrm{dex}$ on its own results in the best-fit age being over-estimated by the fixed value of $0.4\ \mathrm{Gyr}$ (inferred from the overall shift in the spread of best-fit ages), in addition to inflating the formal error by a factor of $\sim 4-5$ (inferred from the increased width of the spread). In many cases, the systematic errors must be interpreted as upper limits, since the tests assume that no measures are taken to suppress them (i.e. the ``naive'' analysis from Section~\ref{sec:analysis} is used).}

\subsection{Multiple populations} \label{sec:systematics_mpops}

The initial set of simulations presented in Section~\ref{sec:simulation} was based on the assumption that all members of 47\,Tuc have the same chemical composition that can be precisely estimated from spectroscopic observations of bright members. However, this assumption is far removed from reality, since globular clusters are known to display significant member-to-member variations of light element abundances (see the review in \citealt{mPOPs_review_main}). This phenomenon is known as the issue of \textit{multiple populations} \citep{multiple_populations,multiple_populations_2}. \annotation{This sentence adds a reference to the first detection of Na-O anti-correlation and two examples of oxygen-driven photometric spread detected with HST and JWST that were suggested by the referee. As discussed in the original response to the referee's comments, I do not believe that this largely methodological section of the manuscript is an appropriate space for a thorough review of the phenomenon of multiple populations.} \revision{Variations in the oxygen abundance \citep{O_Na,Milone_oxygen,marino_47tuc_BD} are most relevant to this study, as $[\mathrm{O/Fe}]$ determines the depth of infrared $\mathrm{H_2 O}$ bands, which dominate the spectral energy distribution at the infrared wavelengths observed with JWST NIRCam.}

\annotation{The referee raised a concern that my analysis does not account for systematic errors due to alpha-enhancement as well as individual light elements other than oxygen. The following two paragraphs refer to my previous study, Gerasimov et al. 2024, that demonstrates that the effects of all individual element abundances on the JWST ultra-wide bands are far subdominant to those of oxygen and carbon, and the effect of carbon only dominates at the lowest stellar masses where atmospheric methane is produced in appreciable amounts.}

\revision{The effects of other light elements on the luminosity function may also be important in certain regimes. Figure~5 of \citet{roman_47Tuc} evaluates the contributions of $11$ individual elements to the observed \texttt{F150W2}$-$\texttt{F322W2} color of the star as a function of stellar mass, which I use as a proxy for the overall impact of the element on the atmosphere and, therefore, the luminosity function of the cluster. The effect of oxygen dominates over the effects of all other elements down to $\sim 0.075\ \mathrm{M}_\odot$ ($T_\mathrm{eff}\sim 1500\ \mathrm{K}$), with the effect of the next most important element (carbon) being weaker by a factor of $\gtrsim 4$, and the effects of other light elements influenced by multiple populations ($\mathrm{N}$, $\mathrm{Na}$, $\mathrm{Mg}$, $\mathrm{Al}$) being even smaller than the effect of carbon. At cooler effective temperatures, the chemical equilibrium of the atmosphere transitions from being oxygen-dominated to being carbon-dominated, which is most apparent in the preferred production of $\mathrm{CH_4}$ \citep{CO_ratio}. In this extremely low temperature regime, the photometric impact of carbon dominates over that of oxygen, while the effect of all other elements remains small. The age error due to the spread of carbon is expected to be suppressed by the fact that most observable brown dwarfs are likely to be located above the point of transition, especially in more metal-poor globular clusters (see e.g. Figure~4 of \citealt{SANDee}).}

\revision{Since there are no publicly available model grids that allow for independent variations in both carbon and oxygen abundances in the metal-poor brown dwarf regime, I only account for variations in oxygen (which is included in the $[\mathrm{\alpha/Fe}]$ parameter of \texttt{SANDee} models) in the simulations described in this section. The effect of carbon may also contribute to the error budget, especially for closer and more metal-rich clusters, and must be investigated with new stellar models in a follow-up study. Since only one element (carbon or oxygen) dominates the offsets in the \texttt{F150W2}$-$\texttt{F322W2} color in all temperature regimes, any degeneracy between light element abundances and age of the cluster may be potentially alleviated by taking into account the color of stars; however, this would inevitably increase the formal error due to the added degree of freedom, and introduce additional errors e.g. due to interstellar reddening.}

To estimate the effect of multiple populations on measured ages, I generated two sets of simulated observations. The first ``\textit{fixed}'' set includes the effect of multiple populations by allowing the simulated stars to have different $[\mathrm{\alpha/Fe}]$ (and, hence, different $[\mathrm{O/Fe}]$), drawn from a skew-normal distribution\footnote{Specifically, I adopted the definition of the skew-normal distribution from \citet{skewnorm}, and converted the chosen Fisher-Pearson coefficient ($-0.43$) into the skewness parameter $\lambda\approx-1.916$ using numerical minimization.} with the mean value of $0.174\ \mathrm{dex}$, the standard deviation of $0.215\ \mathrm{dex}$, and the Fisher-Pearson skewness coefficient of $-0.43$ (i.e., the $\alpha$-poor tail of the distribution is more extended than the $\alpha$-rich tail). These parameters were adopted from the analysis in \citet{roman_47Tuc} of the spectroscopic oxygen abundances in 47\,Tuc from \citet{nominal_C14,nominal_T14}. The distribution of element abundances was assumed to be independent of the stellar mass. Since the distribution of $[\mathrm{\alpha/Fe}]$ is wider than the range of \texttt{SANDee} models at the metallicity of 47\,Tuc, I linearly extrapolated the mass-luminosity relationships beyond the model range.

\begin{figure}[h]
\centering
\includegraphics[width=0.7\textwidth]{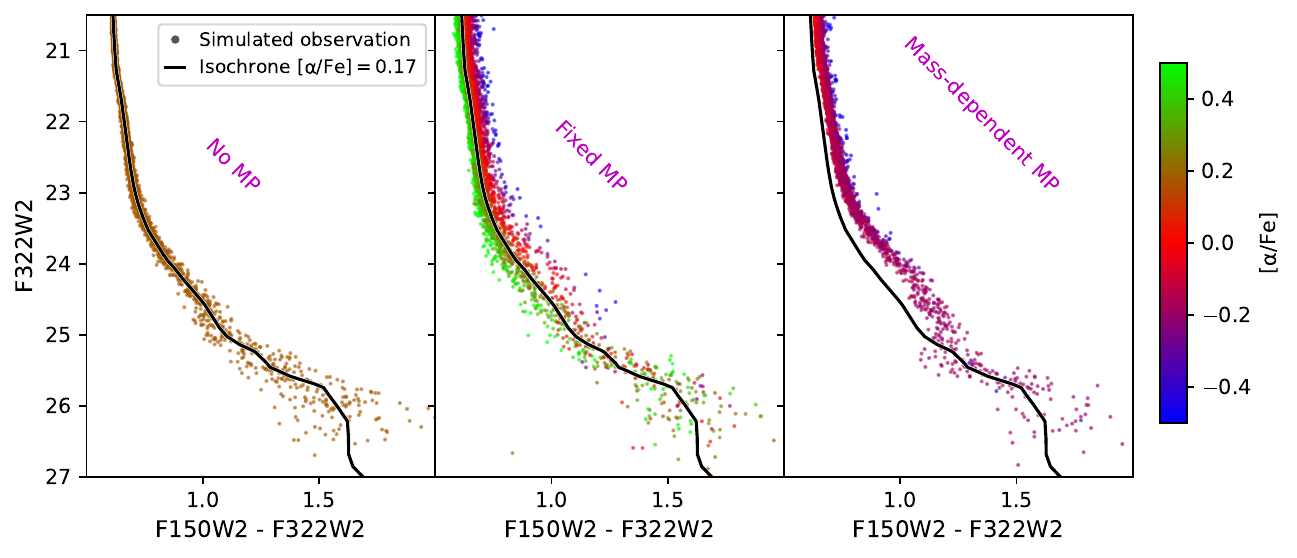}
\caption{Examples of color-magnitude diagrams from simulated observations of 47\,Tuc with JWST NIRCam. \textit{Left:} simulation without multiple populations (MP). All members are assumed to have the mean $\alpha$-enhancement, $[\mathrm{\alpha/Fe}]\approx0.17$. \textit{Middle:} ``\textit{fixed}'' simulation, where the distribution of $[\mathrm{\alpha/Fe}]$ is set to match the spectroscopic measurements of $[\mathrm{O/Fe}]$ in bright members. \textit{Right:} ``\textit{mass-dependent}'' simulation, where the distribution of $[\mathrm{\alpha/Fe}]$ varies with stellar mass according to the constraints from \citet{michele_47Tuc}. The simulations shown in this figure assume two $\mathrm{1\ \mathrm{hr}}$ exposures taken with the temporal baseline of $5\ \mathrm{yr}$. In these examples, $\mathcal{N}=1500$ is used for clarity; however, the actual tests of systemematic errors were carried out for $\mathcal{N}=15,000$.}
\label{fig:mpops}
\end{figure}

\annotation{The following sentence cites previous studies that explored a possible drift in the abundance spread with mass near the end of MS, as requested by the referee.} \revision{Recent JWST observations of 47\,Tuc \citep{marino_47tuc_BD,michele_47Tuc} have unveiled a discontinuity in the color-magnitude diagram of this cluster, which \citet{michele_47Tuc} interpreted as evidence of over-production of $\mathrm{CH_4}$, suggesting that the distribution of} $[\mathrm{O/Fe}]$ may vary with stellar mass near the hydrogen-burning limit \revisionsecond{(see also an earlier claim that the distribution of light element abundances may have mass dependence in \citealt{early_O_mass_claims})}. In particular, \citet{michele_47Tuc} estimated that the upper limit of the oxygen distribution in 47\,Tuc is reduced from $\sim0.5\ \mathrm{dex}$ to $\sim-0.1\ \mathrm{dex}$ as the stellar mass is lowered from $0.6\ \mathrm{M}_\odot$ to $0.08\ \mathrm{M}_\odot$. I implemented this effect in the second ``\textit{mass-dependent}'' ($M$-dep) set of simulations, where the distribution of $\alpha$-enhancement in the cluster is rescaled with stellar mass to meet the constraints from \citet{michele_47Tuc}.

Examples of simulated observations from both fixed and mass-dependent sets, as well as a simulation without multiple populations, are shown in Figure~\ref{fig:mpops}. The scatter in the inferred cluster age, and the modal offset from the true age are shown in Figure~\ref{fig:systematic} with red markers. The presence of multiple populations increases the spread in age by factors of $\sim2$ and $\sim3$ in the fixed and mass-dependent cases, respectively. The latter case also introduces a large systematic bias of $\approx 0.45\ \mathrm{Gyr}$ towards older ages, since the reduced oxygen abundance in brown dwarf members reddens their colors and, therefore, makes them appear older. The systematic offset may be suppressed by using an isochrone with mass-dependent oxygen abundance, fitted to the color-magnitude diagram of the cluster.

\subsection{Unresolved binaries} \label{sec:systematics_binaries}

Unresolved binary members of globular clusters appear up to $\approx0.75\ \mathrm{mag}$ brighter and, for binary systems with unequal masses, redder than the isochrone predictions for single stars. The corresponding redistribution of the luminosity function may impose a significant systematic error on the inferred cluster age, unless the effect of binary stars is included in the model. The population of binary stars in a globular cluster can be parameterized in terms of the overall fraction of binary members, $X$, and the distribution of companion mass ratios, $q=M_s/M_p$, where $M_p$ is the mass of the primary component, $M_s$ is the mass of the companion, and $q\leq1$. The distribution of $q$ is often approximated as a power law (see \citealt{pairing_algorithm} for a review of pairing algorithms), $P(q)\propto q^\beta$, where $\beta$ is the power law slope. Positive $\beta$ values favor equal mass systems (the ``\textit{twin peak}'' effect, \citealt{twin_peak,twin_peak_2}).

While the binary fraction of field stars may exceed $X=50\%$ \citep{solar_binary_fraction}, the binary fractions of globular clusters tend to be much lower ($X\lesssim 10\%$, \citealt{GC_binary_fraction}), likely due to the characteristically old ages of globular clusters and dissociation of binary systems over time \citep{GC_binary_age_evolution_1,GC_binary_age_evolution_2}. For 47\,Tuc, the available estimates of $X$ include measurements from the color-magnitude diagram ($3\%<X<12\%$, \citealt{GC_binary_fraction}), radial velocities ($X=2\%\pm1\%$, \citealt{binary_fraction_47Tuc_RV}), and eclipsing binaries ($X=13\%\pm6\%$, \citealt{binary_fraction_47Tuc_EB}). The largest binary fractions of $X\gtrsim50\%$ have been identified in globular clusters associated with the Sagittarius Stream (Terzan~7, Palomar~12, Arp~2; \citealt{Sagittarius_GCs,GC_binary_age_evolution_2,more_Sollima_binaries}). The binary fraction is also expected to be lower in the brown dwarf regime compared to the main sequence population of the cluster \citep{2007prpl.conf..427B,2018MNRAS.479.2702F}, although it must be noted that this conclusion is based on studies of field objects, as no brown dwarf binaries in globular clusters have been identified so far.

To estimate the systematic errors in inferred ages due to unaccounted binary members, I generated $18$ sets of simulations for all combinations of $X\in\{1,2,5,10,20,50\}\%$ and $\beta\in\{-0.5,1,4\}$ \annotation{The referee mentioned in their report that my analysis only incorporated bottom-heavy mass ratio distributions; however, this section explores a wide range of $\beta$ from extremely bottom-heavy (which is the worst case scenario) to nearly flat}. In the simulations, I randomly assigned a fraction of members to be unresolved binaries according to the prescribed value of $X$, and drew their mass ratios, $q$ from the power law distribution with the prescribed slope $\beta$. The distribution of $q$ was confined to the range $q\in[0.01,1]$. The observed magnitudes of the binary systems were then updated according to the following equation:

\begin{equation}
    m=-2.5\log_{10}\left(10^{-0.4\ m_p} + 10^{-0.4\ m_s}\right)
    \label{eq:binaries}
\end{equation}

\noindent where $m_p$ and $m_s$ are the magnitudes of the primary star and its companion, given by the theoretical mass-luminosity relationship at their corresponding masses. I did not account for triple and higher-order multiple star systems in any of the simulations, as the frequency of their occurrence is expected to be very small \citep{triple_stars}.

The spread in the inferred cluster ages for these simulations and the systematic offsets are shown in Figure~\ref{fig:systematic} with green markers. The cases with $X=50\%$ are not shown in the figure, since the associated errors are very large and exceed the plot limits. For $X\leq10\%$, the contribution of unresolved binaries to both spread and offset remain below $0.1\ \mathrm{Gyr}$ with the exception of the case $(X,\beta)=(10\%,4)$, for which the spread is increased by a factor of $2-3$. For the vast majority of cases with $X\geq20\%$, the added errors in best-fit cluster age exceed $0.2\ \mathrm{Gyr}$, reaching as high as $1.2\ \mathrm{Gyr}$ in the worst-case scenario ($X=50\%$, $\beta=4$). The inferred ages are underestimated, since binary brown dwarfs are brighter and, therefore, appear younger than their true age. I note that binary fractions over $20\%$ are, in general, clearly visible in the color-magnitude diagram, and the induced systematic errors may be reduced by accounting for binary systems in the likelihood model.

\subsection{Modeling errors} \label{sec:systematics_modelling}

For a real NIRCam observation of a globular cluster, the photometric completeness is typically estimated by injecting artificial stars with a realistic PSF into the image and re-running the astro/photometric pipeline to determine the fraction of recovered artificial stars as a function of magnitude (e.g., \citealt{rolly_6397,michele_47Tuc}). The true completeness may be offset from the estimated values due to imperfect PSF modeling, misidentification of field contaminants as cluster members and vice versa, image artifacts, small number statistics at faint magnitudes and other effects. Furthermore, small deviations in the mass function from the assumed broken power law relationship may be modeled as errors in completeness as well, as they increase or decrease the expected number of cluster members at a given magnitude compared to the power law model prediction.

To quantify this effect, I generated three sets simulated observations with random Gaussian offsets in true completeness of $1\%$, $5\%$ and $10\%$. The offsets were only applied to magnitudes with non-zero true completeness, and they were truncated to ensure $0\leq\mathrm{comp}(m)\leq 1$. I then carried out the likelihood maximization analysis on each simulated observation using the original (unperturbed) completeness corrections in the luminosity function. The scatter in inferred cluster ages is shown in Figure~\ref{fig:systematic} with black markers. For the cases of $1\%$, $5\%$ and $10\%$ completeness errors, the spread in best-fit age increased by $0.04\ \mathrm{Gyr}$, $0.13\ \mathrm{Gyr}$ and $0.26\ \mathrm{Gyr}$, respectively, compared to the baseline value. This trend may be used to estimate the maximum tolerable error in photometric completeness for a given target precision in age.

As discussed earlier, I estimated the effect of uncertainties in the adopted mass-luminosity relationships by artificially perturbing the metallicity of the globular cluster in the model. I generated four sets of simulated observations with the metallicity of the theoretical mass-luminosity relationships offset by $\pm0.1,\pm0.2$ $\mathrm{dex}$. The altered metallicity changes the opacity of brown dwarf atmospheres and, hence, their cooling rate. The likelihood maximization was then carried out using the nominal (spectroscopic) metallicity of 47\,Tuc ($[\mathrm{Fe/H}]=-0.72$). The results of these tests are shown in Figure~\ref{fig:systematic} with magenta markers. For the metallicity offsets of $\pm0.1$ and $\pm0.2$, the age errors increased by approximately $0.2\ \mathrm{Gyr}$ and $0.5\ \mathrm{Gyr}$, respectively, compared to the baseline case. The $+0.2$ case also shows a modal offset of $\approx 0.4\ \mathrm{Gyr}$ towards older ages. \revision{This is the most significant systematic effect considered in the study. It may be potentially reduced by choosing the best-fitting isochrone to the color-magnitude diagram near the hydrogen-burning limit, following \citet{SANDee}.}

\subsection{Age and mass function effects} \label{sec:systematics_age_MF}

The errors in best-fit ages and mass function parameters shown in Figure~\ref{fig:random} are based on the true (simulated) age of $11.5\ \mathrm{Gyr}$ and the ``universal'' mass function. To quantify the effect of true age on the error in the age measurement, I re-ran the analysis on simulated observations with true ages between $9\ \mathrm{Gyr}$ and $13\ \mathrm{Gyr}$ in $1\ \mathrm{Gyr}$ steps. The spread in best-fit ages for each set of simulated observations is shown in Figure~\ref{fig:systematic} with blue markers. The spread increases from $\approx 0.08\ \mathrm{Gyr}$ at the true age of $9\ \mathrm{Gyr}$ to $0.16\ \mathrm{Gyr}$ at $13\ \mathrm{Gyr}$. This trend is primarily driven by the decrease in the number of detected brown dwarfs at older ages due to cooling.

The effect of true mass function was estimated by generating $27$ sets of simulations for every combination of $\alpha_\mathrm{h}\in\{-2,-1.5,-1,-0.5,0,0.5,1,1.5,2\}$ and $\alpha_\mathrm{l}=\{-0.5,0,0.5\}$. The spread in inferred ages for each mass function is shown in Figure~\ref{fig:systematic} with brown markers. More precise ages are obtained for higher (more bottom-heavy) values of $\alpha_\mathrm{h}$ with the expected errors of $0.08\ \mathrm{Gyr}$ and $0.3\ \mathrm{Gyr}$ at $\alpha_\mathrm{h}=+2$ and $-2$, correspondingly. As with true age, higher values of $\alpha_\mathrm{h}$ produce more brown dwarfs near the hydrogen-burning limit, allowing for a tighter constraint on the age of the cluster. On the other hand, the true value of $\alpha_\mathrm{l}$ does not have a noticeable effect due to the narrow range of observable masses below the power law break.

\section{Other globular clusters} \label{sec:other_gcs}

The results in Section~\ref{sec:analysis} were derived for the known properties of the globular cluster 47\,Tuc. While 47\,Tuc is an excellent candidate for this analysis due to its relative proximity and size, the dating method proposed in this study can be applied to the larger population of nearby globular clusters. In this section, I generalize the results of Section~\ref{sec:analysis} to other globular clusters, for which the brown dwarf cooling sequences may be observed with JWST. Since it is computationally expensive and impractical to generate simulated observations for each globular cluster, I instead propose a set of scaling relationships that may be used to transform the age errors of 47\,Tuc to other clusters.

\begin{figure}[h]
\centering
\includegraphics[width=0.7\textwidth]{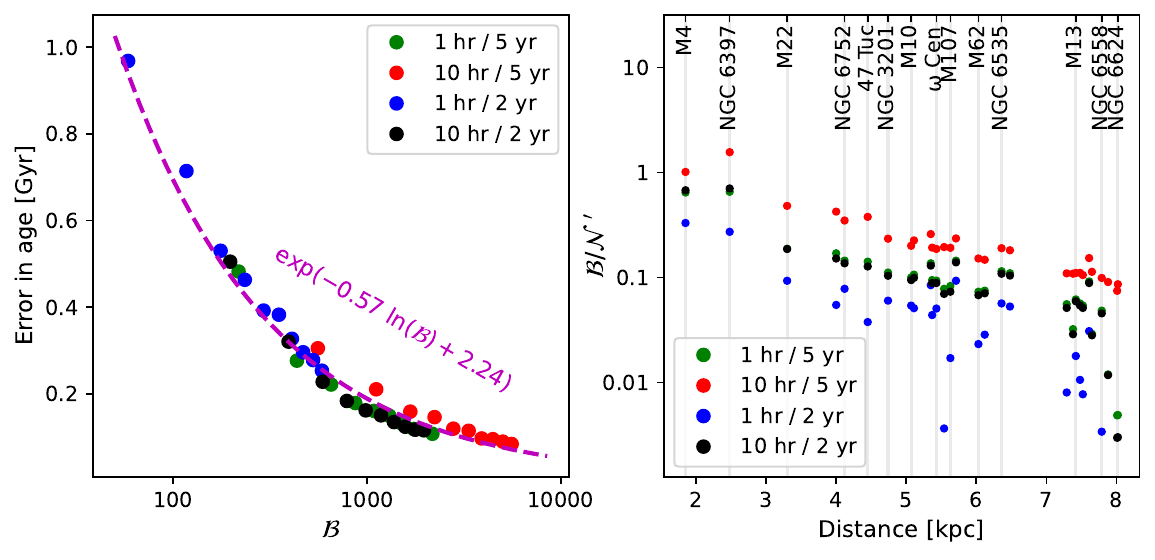}
\caption{\textit{Left:} relationship between the number of observed age-sensitive objects (brown dwarfs and low-mass stars immediately above the hydrogen-burning limit), $\mathcal{B}$, and the expected \revision{formal} error in inferred cluster age. The color-coded observing strategies are the same as in Figure~\ref{fig:random}. The best-fit exponential fit is shown in magenta. \textit{Right:} the ratio of $\mathcal{B}$ and the scaled richness value, $\mathcal{N}'$ for $30$ nearby globular clusters. Selected globular clusters are highlighted with vertical lines and labeled.}
\label{fig:other}
\end{figure}

The left panel of Figure~\ref{fig:other} shows the expected \revision{\textit{formal}} error in the inferred age of 47\,Tuc as a function of parameter $\mathcal{B}$, which I define as the total number of detected objects with stellar masses $M<M_\mathrm{HBL}+\delta M$. Here $M_\mathrm{HBL}$ is the stellar mass at the hydrogen-burning limit ($\approx0.0775\ \mathrm{M}_\odot$ for 47\,Tuc from the \texttt{SANDee} models) and $\delta M$ is an offset that produces the best correlation between the error in age and $\mathcal{B}$. For 47\,Tuc, I estimated $\delta M$ to be $\approx0.0013\ \mathrm{M}_\odot$. The values shown in the left panel of Figure~\ref{fig:other} are a combination of all results from Section~\ref{sec:analysis}, including all $4$ observing strategies and all $10$ considered values of $\mathcal{N}$.

The tight correlation between $\mathcal{B}$ and the expected error in age suggests that $\mathcal{B}$ may be used as a predictor of the latter. To first order, $\mathcal{B}$ represents the number of detected brown dwarfs in a given observation. The small correction, $\delta M$, is introduced to account for low-mass stars just above the hydrogen-burning limit, as they may also be sensitive to age. I determined the best-fit relationship between age error and $\mathcal{B}$ to be the equation below, and assumed that the same relationship holds for all globular clusters.

\begin{equation}
    \mathrm{formal\ age\ error}\ [\mathrm{Gyr}]=e^{-0.57 \ln(\mathcal{B}) + 2.24}
    \label{eq:B_to_error}
\end{equation}

The right panel of Figure~\ref{fig:other} shows the ratio of $\mathcal{B}$ to $\mathcal{N}'$ for $30$ nearest globular clusters that are listed in both \citet{GC_masses} and \citet{GC_harris}, and are not overly extinguished ($E(B-V)<0.5$). $\mathcal{N}'$ is the richness of field, equivalent to $\mathcal{N}$, but scaled to the color-magnitude diagram of each globular cluster. The original value of $\mathcal{N}$ was defined as the number of 47\,Tuc members in the observed field between \texttt{F322W2} magnitudes $21$ and $22$, which corresponds to the stellar masses of $\approx 0.1855\ \mathrm{M}_\odot$ and $\approx 0.1234\ \mathrm{M}_\odot$, respectively, according to the \texttt{SANDee }mass-luminosity relationship for the chemical composition of 47\,Tuc. The scaled richness, $\mathcal{N}'$, is defined as the number of cluster members in the observed field between the same stellar masses, but with the magnitude range that corresponds to those masses according to the \texttt{SANDee} mass-luminosity relationship for the metallicity of the cluster, adopted from \citet{GC_masses}. For 47\,Tuc, $\mathcal{N}'=\mathcal{N}$.

The values of $\mathcal{B}/\mathcal{N}'$ in the right panel of Figure~\ref{fig:other} were calculated by interpolating the \texttt{SANDee} models to the metallicity of the cluster from \citet{GC_masses} and interstellar reddening from \citet{GC_harris}, calculating the luminosity function using Equation~\ref{eq:MF_to_LF}, applying the photometric completeness to the luminosity function and integrating the luminosity function in the appropriate magnitude ranges to obtain the ratio of $\mathcal{B}$ and $\mathcal{N}'$. In each case, the age of $11.5\ \mathrm{Gyr}$ and the ``universal'' mass function were used. The right panel of Figure~\ref{fig:other} shows that $\mathcal{B}/\mathcal{N}'$ generally decreases with distance, as fewer brown dwarfs fall within the sensitivity of JWST in more distant clusters. However, metallicities and interstellar reddening along the line of sight introduce small corrections to the overall trend. Notably, more brown dwarfs can be detected in NGC~6397 than in M4 for some of the observing strategies despite the larger distance to NGC~6397. This is because NGC~6397 is more metal-poor and far less extinguished.

The calculated values of $\mathcal{B}/\mathcal{N}'$, as well as the range of \texttt{F322W2} magnitudes used in calculating $\mathcal{N}'$ are listed in Table~\ref{tab:scaling}. The table can be used as a lookup reference to predict the expected \revision{formal} age error for a given globular cluster and observing strategy. The example below demonstrates this calculation for the globular cluster NGC~6397:

\begin{enumerate}
    \item We start by choosing the observing strategy and looking up the appropriate $\mathcal{B}/\mathcal{N}'$ ratio for NGC~6397. For this example I chose two $1\ \mathrm{hr}$ exposures $5\ \mathrm{yr}$ apart, which corresponds to $\mathcal{B}/\mathcal{N}'=0.66$.
    \item Next, we must set the intended scaled richness of the observation, $\mathcal{N}'$, which for NGC~6397 corresponds to the total number of members with \texttt{F322W2} magnitudes between $20.1$ and $20.6$. The NIRCam imaging of NGC~6397 in \citet{rolly_6397} yielded $\sim 85$ members in the required magnitude range. Assuming that the planned JWST program has a similar scope, we can set $\mathcal{N}'=85$. \revision{Note that this value of $\mathcal{N}'$ is over an order of magnitude smaller compared to, e.g., the example of 47\,Tuc considered earlier. We therefore expect large formal errors.}
    \item It is now possible to calculate $\mathcal{B}=85\times0.66\approx 56$ for the chosen observing strategy. \revision{We therefore expect each observation of NGC~6397 comparable to that in \citet{rolly_6397} to reveal $\sim56$ age-sensitive objects (mostly brown dwarfs).}
    \item Finally, we substitute the value of $\mathcal{B}$ into Equation~\ref{eq:B_to_error} to estimate the expected age error as $\approx0.95\ \mathrm{Gyr}$.
    \item If, instead, a precision of $0.3\ \mathrm{Gyr}$ is required, we can use Equation~\ref{eq:B_to_error} to estimate the required value of $\mathcal{B}$ as $\sim 420$.
    \item Since the required value of $\mathcal{B}$ is now $\sim7$ times larger than the expected value for a single observation, approximately $7$ fields comparable to that in \citet{rolly_6397} must be observed in order to attain the required precision.
\end{enumerate}

\begin{table}[h]
\centering
\caption{Lookup table for estimating the expected \revision{formal} errors in best-fit age estimates for nearby globular clusters.}\label{tab:scaling}%
\begin{tabular}{lc cccc c}
\toprule
Cluster & Distance & \multicolumn{4}{c}{$\mathcal{B}/\mathcal{N}'$} & Magnitude range \\
\cmidrule(lr){3-6}
 & [$\mathrm{kpc}$] & $1\ \mathrm{hr}$ / $5\ \mathrm{yr}$ & $10\ \mathrm{hr}$ / $5\ \mathrm{yr}$ & $1\ \mathrm{hr}$ / $2\ \mathrm{yr}$ & $10\ \mathrm{hr}$ / $2\ \mathrm{yr}$ & for $\mathcal{N}'$ \\
\midrule
M4   & 1.8  & 0.64  & 1.01 & 0.33 & 0.68 & $19.1-20.1$ \\
NGC~6397   & 2.5  & 0.66  & 1.56 & 0.27 & 0.70 & $20.1-20.6$ \\
M22   & 3.3  & 0.19  & 0.48 & 0.09 & 0.19 & $20.3-21.3$ \\
M71   & 4.0  & 0.17  & 0.42 & 0.05 & 0.15 & $20.8-21.8$ \\
NGC~6752   & 4.1  & 0.14  & 0.35 & 0.08 & 0.14 & $20.7-21.7$ \\
47~Tuc   & 4.5  & 0.14  & 0.38 & 0.04 & 0.13 & $21.0-22.0$ \\
NGC~3201   & 4.7  & 0.11  & 0.23 & 0.06 & 0.10 & $21.1-22.0$ \\
M10   & 5.1  & 0.10  & 0.20 & 0.05 & 0.09 & $21.2-22.2$ \\
M12   & 5.1  & 0.11  & 0.23 & 0.05 & 0.10 & $21.2-22.2$ \\
M55   & 5.4  & 0.14  & 0.26 & 0.08 & 0.13 & $21.6-22.2$ \\
M28   & 5.4  & 0.09  & 0.19 & 0.04 & 0.09 & $21.4-22.4$ \\
$\mathrm{\omega}$~Cen   & 5.4  & 0.09  & 0.19 & 0.05 & 0.09 & $21.3-22.3$ \\
NGC~6352   & 5.5  & 0.08  & 0.19 & 0.00 & 0.07 & $21.6-22.6$ \\
M107   & 5.6  & 0.08  & 0.19 & 0.02 & 0.07 & $21.5-22.6$ \\
NGC~4372   & 5.7  & 0.14  & 0.24 & 0.09 & 0.14 & $21.9-22.4$ \\
M62   & 6.0  & 0.07  & 0.15 & 0.02 & 0.07 & $21.7-22.7$ \\
ESO~456-78   & 6.1  & 0.08  & 0.15 & 0.03 & 0.07 & $21.6-22.7$ \\
NGC~6535   & 6.4  & 0.11  & 0.19 & 0.06 & 0.11 & $22.0-22.7$ \\
NGC~4833   & 6.5  & 0.11  & 0.18 & 0.05 & 0.10 & $22.1-22.7$ \\
NGC~6522   & 7.3  & 0.06  & 0.11 & 0.01 & 0.05 & $22.1-23.1$ \\
NGC~6712   & 7.4  & 0.03  & 0.11 & 0.00 & 0.03 & $22.2-23.2$ \\
M13   & 7.4  & 0.06  & 0.11 & 0.02 & 0.06 & $22.0-23.0$ \\
M5   & 7.5  & 0.06  & 0.11 & 0.01 & 0.05 & $22.0-23.1$ \\
NGC~6717   & 7.5  & 0.05  & 0.11 & 0.01 & 0.05 & $22.1-23.1$ \\
NGC~6541   & 7.6  & 0.09  & 0.15 & 0.03 & 0.09 & $22.4-23.0$ \\
NGC~6362   & 7.7  & 0.03  & 0.11 & 0.00 & 0.03 & $22.2-23.2$ \\
NGC~6558   & 7.8  & 0.05  & 0.10 & 0.00 & 0.05 & $22.2-23.2$ \\
E3   & 7.9  & 0.01  & 0.09 & 0.00 & 0.01 & $22.3-23.3$ \\
NGC~6342   & 8.0  & 0.00  & 0.07 & 0.00 & 0.00 & $22.4-23.4$ \\
NGC~6624   & 8.0  & 0.00  & 0.09 & 0.00 & 0.00 & $22.4-23.4$ \\
\end{tabular}
\end{table}

\section{Conclusion}\label{sec:conclusion}

Globular clusters are prime targets for galactic archaeology, as their properties are indicative of the evolutionary history of the Milky Way. The ages of globular clusters are a key diagnostic of their natal environments. In this study, I presented a new approach to dating globular clusters that relies on the cooling behavior of their substellar members. This approach has a distinct set of systematic errors from all other dating methods, and may therefore both alleviate the existing tension in globular cluster ages from different techniques, and provide an independent test of the state-of-the-art stellar models.

The new dating method requires a large sample of brown dwarf members with photometric magnitudes. Due to the faint luminosities and infrared colors of brown dwarfs, this type of observation can only be carried out with JWST at present. In this study, I specifically focused on the \texttt{F150W2} and \texttt{F322W2} bands of NIRCam, since their wide wavelength coverage is most suitable for faint and cool objects.

The specific procedure for measuring the age of a globular cluster is laid out in Section~\ref{sec:analysis}. It involves imaging the same part of the globular cluster twice several years apart, selecting \textit{bona fide} cluster members with proper motion filtering and determining the most compatible age and mass function through likelihood maximization. My proposed method operates directly on the measured magnitudes and, unlike most existing approaches to luminosity function analysis, does not require binning of data.

I tested the new dating technique on a large set of simulated observations of the globular cluster 47\,Tuc. 47\,Tuc is an ideal target for this analysis due to its large mass and close distance. Furthermore, 47\,Tuc has already been targeted with NIRCam, albeit without a sufficiently long temporal baseline to unambiguously identify the substellar sequence. I took a recent observation of 47\,Tuc with JWST, analyzed in \citet{michele_47Tuc}, to represent a typical JWST field in this cluster. The specific findings of this study from the simulated observations of 47\,Tuc are as follows:

\begin{enumerate}
    \item For a single JWST field similar to the one in \citet{michele_47Tuc}, the expected \revision{formal} errors in the inferred cluster age vary between $\approx0.3$ and $\approx1\ \mathrm{Gyr}$ depending on the observing strategy. The lowest error of $\approx0.3\ \mathrm{Gyr}$ is obtained for two $10\ \mathrm{hr}$ exposures taken $5\ \mathrm{yr}$ apart. The highest error of $\approx 1\ \mathrm{Gyr}$ is obtained for two $1\ \mathrm{hr}$ exposures taken $2\ \mathrm{yr}$. The other two observing strategies considered in this study -- $1\ \mathrm{hr}$ exposures $5\ \mathrm{yr}$ apart and $10\ \mathrm{hr}$ exposures $2\ \mathrm{yr}$ apart -- yield nearly the same formal age error of $\approx 0.5\ \mathrm{Gyr}$. Based on these results, taking two $1\ \mathrm{hr}$ exposures $5\ \mathrm{yr}$ apart appears to be the most efficient use of JWST time.
    \item For all observing strategies except for the least demanding option ($1\ \mathrm{hr}$ exposures $2\ \mathrm{yr}$ apart), \revision{formal} errors as low as $0.2\ \mathrm{dex}$ are attainable by observing $3$ or $4$ non-overlapping fields in the cluster.
    \item All observing strategies also provide constraints on the power law slope of the cluster mass function. \revision{Formal} errors in the power law slope below $0.1$ are well within the sensitivity limit. However, if the mass function changes its slope near the hydrogen-burning limit, as it does in the ``universal'' mass function \citep{Kroupa_IMF}, the low-mass slope will likely remain unconstrained. Fortunately, due to the small range of masses on the brown dwarf cooling sequence, even large uncertainties in the mass function do not significantly impact the age estimate.
    \item The estimated errors above were based on the true age of $11.5\ \mathrm{Gyr}$ and the ``universal'' mass function. The expected error in best-fit age increases for older true ages and less bottom-heavy mass functions, since both effects decrease the number of detected brown dwarfs. In particular, the \revision{formal} error increases by a factor of $2$ when the true age is increased from $9\ \mathrm{Gyr}$ to $13\ \mathrm{Gyr}$, and by a factor of $3$ when the mass function slope is decreased from $+2$ to $-2$ ($\alpha_\mathrm{h}$ in Equation~\ref{eq:powerlaw}).
\end{enumerate}

\revision{As with other dating techniques, the error budget is likely to be dominated not by formal errors, but by systematic effects.} In Section~\ref{sec:systematics}, I carried out a detailed analysis of systematic errors associated with the new dating technique, as they are likely to dominate the overall error budget. My investigation of systematic errors may be summarized as follows:

\begin{enumerate}
    \item The presence of multiple populations in the cluster may introduce systematic offsets as large as $0.5\ \mathrm{Gyr}$, and inflate the formal error by a factor of $2-3$ if no attempt to correct for this effect is made. This systematic error is at its largest if the distribution of light element abundances varies significantly with stellar mass, as has been suggested by \citet{michele_47Tuc}. It is however unclear how prevalent this effect is in other globular clusters. The systematic error may be reduced if the mass dependence of the chemical spread is included in the adopted mass-luminosity relationships.
    \item The presence of unresolved binaries may be ignored for nearly all globular clusters, as binary fractions under $10\%$ and bottom-heavy distributions (i.e. the power-law slope, $\beta<0$) of companion masses appear common \citep{GC_binary_fraction}. However, for globular clusters with binary fractions over $20\%$, such as the clusters of the Sagittarius stream, the effect must be accounted for in the likelihood model. This may be non-straightforward, since the binary fraction and companion mass distribution may be different in the brown dwarf regime compared to the main sequence.
    \item Systematic errors due to an offset in the metallicity adopted in stellar models and, hence, inaccurate brown dwarf cooling rates may inflate formal errors by up to a factor of $5$ for the specific scenarios considered in this study. These errors may be reduced by leveraging the color-magnitude diagram of the brown dwarf cooling sequence to select the appropriate cooling curve during the analysis.
\end{enumerate}

\annotation{In the added paragraph below, I estimated the approximate final error budget to be of the order of $1.5$ - $2$ Gyr, which should be interpreted as the worst case scenario, as it assumes that nothing is done to reduce those errors. The error budget is however clearly far larger than the formal errors from Section 3.}

\revision{The two most important contributions to the error budget according to Figure~\ref{fig:systematic} are multiple populations (particularly when the chemical spread changes at lower stellar masses) and metallicity offset (this effect is likely larger for 47\,Tuc than for most globular clusters due to its higher metallicity). In isolation, these two effects may inflate the formal errors by factors of $\sim3$ and $\sim5$, respectively, if no measures are taken to reduce them. For the nominal formal error in age of $0.3\ \mathrm{Gyr}$, these factors correspond to the overall error budget of $\sim 1.5-2\ \mathrm{Gyr}$ (all effects added in quadrature), which is comparable to, e.g., systematic error budgets of the MSTO-based age measurements in \citet{GC_ages_2}; however, the systematic effects in MSTO-based ages and brown dwarf-based ages are almost completely independent. As with MSTO measurements, the error budget can be reduced (potentially significantly) with more sophisticated analysis.}

While the simulated observations in this study were primarily aimed at the known parameters of 47\,Tuc, the new dating technique may be applied to other nearby globular clusters. In Section~\ref{sec:other_gcs}, I derived a set of scaling relationships that may be used to convert the calculated age errors for any of the $30$ globular clusters listed in Table~\ref{tab:scaling}.

%----------------------------------------------------------

\printbibliography

%----------------------------------------------------------

\end{document}